\documentclass[onecolumn]{aastex631}

\def\fer{{{\it Fermi}}\/}

\def\aap{A\&A}

\def\apjl{ApJL}
\def\apjs{ApJS}

\shorttitle{Identifying the 3FHL catalog: Gemini results}
\shortauthors{Rajagopal et al.}

\usepackage{natbib}
\usepackage{float}
\usepackage{color}
\usepackage{graphicx}
\usepackage{textcomp,gensymb}
\usepackage{array}
\usepackage{mathtools}
\usepackage{enumitem}
\usepackage{longtable}
\usepackage{hyperref}
\DeclareUnicodeCharacter{2212}{-}

\begin{document}
\title{IDENTIFYING THE 3FHL CATALOG: VI. RESULTS OF THE 2019 GEMINI OPTICAL SPECTROSCOPY CAMPAIGN}

\author[0000-0002-8979-5254]{M. Rajagopal}
\affiliation{Department of Physics and Astronomy, Clemson University, Clemson, SC 29634, USA}
\author[0000-0002-8472-3649]{L. Marcotulli}
\altaffiliation{NHFP Einsten Fellow}
\affil{Department of Physics, Yale University, 52 Hillhouse Avenue, New Haven, CT 06511, USA}
\author[0000-0002-6633-7891]{K. Labrie}
\affiliation{Gemini Observatory, NSF's NOIRLab, 670 N. A’ohoku Place, Hilo, Hawai‘i 96720, USA}
\author{S. Marchesi}
\affiliation{INAF - Osservatorio di Astrofisica e Scienza dello Spazio di Bologna, Italy}
\affiliation{Department of Physics and Astronomy, Clemson University, Clemson, SC 29634, USA}
\author[0000-0002-6584-1703]{M. Ajello} 
\affiliation{Department of Physics and Astronomy, Clemson University, Clemson, SC 29634, USA}

\begin{abstract}
Active galactic Nuclei (AGNs) with their relativistic jets pointed towards the observer, are a class of luminous gamma-ray sources commonly known as blazars. The study of 
this source class is essential to unveil the physical processes powering these extreme jets, 
to understand their cosmic evolution, as well as to indirectly probe the extragalactic background light. To do so, however, one  needs to correctly classify and measure a redshift for a large sample of these sources. The Third {\it Fermi}--LAT Catalog of High-Energy Sources (3FHL) contains $1212$ blazars detected at energies greater than $10$\,GeV. However, $\sim$25\% of these sources are unclassified and $\sim$56\% lack redshift information. To increase the optical completeness of blazars in the 3FHL catalog, we devised an optical spectroscopic follow up campaign using 4\,$m$ and 8\,$m$ telescopes. In this paper, we present the results of the last part of this campaign, where we observed 5 blazars using the 8\,$m$ Gemini-S telescope in Chile. We report all the 5 sources to be classified as BL Lacs, a redshift lower limit for 2 sources, and featureless spectra for the remaining 3 sources. We also performed a one-zone leptonic fit to the two sources with the redshift lower limits.
\end{abstract}

\section{Introduction}

Blazars are a peculiar class of active galactic nuclei (AGNs) which dominate the observable $\gamma$-ray Universe \citep{1FGL_2010,2FGL_2012,3FGL_2015,4FGL_2022}.  Their emission is a byproduct of non-thermal plasma traveling relativistically towards the observer along the jet magnetic field lines.
The acceleration of charged particles in the presence of a magnetic field generates synchrotron emission, observed as a bump from infrared to X-rays in the spectral energy distribution (SED) of the blazars. When the low energy photons in the medium are upscattered via inverse Compton process by the same particles, it leads to a second bump (from X-rays to $\gamma$-rays) in their SED. Furthermore, the jet peculiar orientation causes order-of-magnitude flux amplification, which enables us to detect them at very high redshifts. Therefore, blazars are extremely valuable sources to understand the physics of AGN jets and their evolution through cosmic time.

Of particular importance are the blazars detected at the highest $\gamma$-ray energies, $E>10\,\rm GeV$. In fact these sources are some of the most powerful accelerators in the universe, being able to accelerate electrons to beyond 100 TeV \citep[e.g.][]{Costamante_2001,Tavecchio_2011}.
Therefore, innovative scientific results can be obtained by studying these extreme blazars, provided their redshift ($z$) is known. In primis,  these sources can help us understand jet emission processes \citep[e.g.][]{ghisellini17} or the cosmological evolution of the class \citep{ajello12, ajello14}.
Furthermore, they are significant in a cosmological context, as they can be used to indirectly probe the extragalactic background light (EBL), i.e.~all the radiation emitted by stars and galaxies and reprocessed radiation from interstellar dust 
\citep{ackermann12,dominguez13, EBL_2018, Deasi_EBL_2019}.
Finally, these sources will enable us to provide a map of targets for the upcoming Cherenkov Telescope Array (CTA, \citealp[e.g.][]{Goldoni_2021}), and discovering possible candidate neutrino emitters \citep{IceCube_2018}.

The Third \fer--LAT Catalog of High-Energy Sources \citep[3FHL][]{ajello17} is the latest catalog of sources emitting photons of $E>10\,\rm GeV$ detected  by the Large Area Telescope (LAT) aboard the {\it Fermi Gamma-ray Space Telescope} \citep{atwood09}. It encompasses seven years of observations and contains more than 1500 sources, the vast majority of which (78\%) are blazars \citep{ajello17}.
Out of these 1212 blazars, only $44\%$ have a redshift measurement \citep{ajello17,EBL_2018}. To overcome this limitation, extensive optical spectroscopic campaigns, targeting those 3FHL objects still lacking redshift and classification, must be performed.

Besides being used for redshift determination, optical spectroscopy campaigns of blazars \citep[e.g.][]{Massaro_2016,deMenezes_2020,Pena-Herazo_2019, Pena-Herazo_2021} are also essential to distinguish between blazar sub-classes.
The standard division between the classes occurs based on the line equivalent width (EW) in the optical spectrum, where flat spectrum radio quasars (FSRQ) have EW$>5$\r{A} while BL Lacertae objects (BL Lac) show EW$<5$\r{A} \citep{urry95,ghisellini17}. The emission lines in the BL Lac spectra are weak or absent and the lines in FSRQs are extremely prominent. 
FSRQs are generally high redshift objects with average luminosity larger than that of BL Lacs \citep{padovani92,paiano17}.
In the 3FHL \citep{ajello17}, 14\% of blazars are of the FSRQ type and 62\% are of the BL Lac type. The blazar sources not classified as FSRQ or BL Lac are listed as blazar candidates of uncertain type (BCU), and constitute $\approx 25\,\%$ of the reported blazar sample.
Obtaining a spectroscopically complete classification of the blazars observed by \fer--LAT in the $\gamma$-ray regime is essential to validate claims of different cosmological evolution of the two classes \citep{ajello12,ajello14}. 

The ground based telescopes used in spectroscopic campaigns are generally of the 4\,$m$, 8\,$m$ and 10\,$m$ class type. While the 10\,$m$ and 8\,$m$ class telescopes are shown to be significantly more effective in obtaining redshift measurements for blazars \citep[60--80\% versus 25--40\% success rate, see, e.g.][]{paiano17,marchesi18}, even 4\,$m$ class telescopes have proven to be useful for effectively distinguishing between the two different blazar subclasses \citep[see ][]{shaw13,massaro14,paggi14,ricci15,landoni15,marchesini16,alvarez16a,alvarez16b,rajagopal2021}.  
This work is part of a larger spectroscopic follow-up campaign started in 2017 to classify the BCUs in the 3FHL catalog and measure their redshift. 
This effort has been divided in a combination of BCU classification based on machine learning algorithms \citep{Kaur_2019, silver2020, Joffre_2022}, as well as optical spectroscopic classification and redshift identification using $4\,m$ and $8\,m$ facilities \citep{marchesi18, desai2019, rajagopal2021}. The major results of this campaign are summarized in  Table~\ref{tab:summary}.
\\
As part of this campaign, in 2019 we carried out a Gemini program\footnote{{\it Fermi}-Gemini proposal approved Program ID: GS-2019A-Q-213, PI: Dr. Stefano Marchesi} to increase the spectroscopic completeness of the 3FHL catalog. The observations were partially fulfilled between 06 February 2019 and 04 June 2019 and 5 targets were observed. The results of this latest effort are presented in this paper which is organized as follows:
Section~\ref{sample_sel} reports the criteria used for our sample selection, Section~\ref{obs} describes the methodology used for the source observation and spectral extraction procedures, Section~\ref{spectral} lists the results of this work, both, for each individual source and also in general terms, while Section~\ref{sedmodels} reports the SED modeling performed on two of the sources (with lower limits on $z$). Section~\ref{conclusion} reports the conclusions inferred from this spectroscopic campaign.

\section{Sample Selection}\label{sample_sel}
We selected the 5 objects in our sample among the BCUs in the 3FHL catalog, using the following criteria:
\begin{itemize}
\item The 3FHL source should be bright in the hard $\gamma$-ray spectral regime ($f_{\rm 50-150 GeV}>$10$^{-12}$ erg s$^{-1}$ cm$^{-2}$). This selection criteria
ensures that the completeness of the 3FHL catalog evolves to lower fluxes as more optical observations are performed.
\item The target should  be observable from Cerro Pachón with an altitude above the horizon $\delta$$>$40\,\degree (i.e., with airmass $<$1.5): this corresponds to a declination range -80\degree $<$ Dec $<$ 20\degree. Furthermore, as we were granted observing time in October, we select targets that could be observable during this month
(i.e., it should have R.A. $\geq$ 09h0m00s and R.A. $\leq$ 0h30m00s).
\item The target should  not have any optical spectrum or classification already reported by works focusing on complementary catalogs \citep[e.g. the 3FGL or the 4FGL,][ and references therein]{Massaro_2016,deMenezes_2020,Pena-Herazo_2019, Pena-Herazo_2021}.
\end{itemize}

The sources used in our sample and their properties are listed in Table~\ref{tab:sample}.
\begingroup
\renewcommand*{\arraystretch}{1.8}
\begin{table}[t!]
\centering
\caption{Summary of the optical spectroscopic campaign to classify the BCUs in the 3FHL. }
%\scalebox{0.64}{
\resizebox{\textwidth}{!}{
\hspace{-6cm}
    \begin{tabular}{c|c c|c c|c c|c c|c c|c c| c c | c c}
                & \multicolumn{2}{c|}{3FHL \citep{ajello17}} & \multicolumn{2}{c|}{\citet{Kaur_2019}} & \multicolumn{2}{c|}{\citet{marchesi18}} & \multicolumn{2}{c|}{\citet{desai2019}} & \multicolumn{2}{c|}{\citet{silver2020}} & \multicolumn{2}{c|}{\citet{rajagopal2021}} & 
                \multicolumn{2}{c|}{\citet{Joffre_2022}} &
                \multicolumn{2}{c}{This work}\\
        \hline
                &  Tot \# & w. $z$ &  Tot \# & w. $z$ &  Tot \# & w. $z$ &  Tot \# & w. $z$ &  Tot \# & w. $z$ &  Tot \# & w. $z$ & Tot \# & wo $z$ & Tot \# & wo $z$ \\
        \hline
        Blazars & 1212 & 536  & 1212      &  536 & 1212      & 543 (+7) & 1212      &  551 (+8) & 1212      & 551 &  1212     & 567 (+16) &  1212     & 567 &   1212       & 569 (+2) \\
        \hline
        FSRQs   & 172  & 163  & 172       &  163 & 173 (+1)  & 164 (+1) & 173       &  164      & 173       & 164 & 176 (+3)  & 166 (+2) & 176 & 166  & 176 & 166       \\
        BL Lacs & 750  & 344  & 786 (+36) &  344 & 813 (+27) & 350 (+7) & 836 (+23) &  358 (+8) & 851 (+15) & 358 & 876 (+25) & 508 (+14) & 896 (+20) & 372 & 901 (+5) & 374 (+2) \\
        BCU     & 290  & 29   & 254 (-36) &  29  & 226 (-28) & 29       & 203 (-23) &  175      & 188 (-15) & 29  & 160 (-25) & 132  & 140 (-20) & 29 & 135 (-5) & 29      \\
    \end{tabular}}
    \label{tab:summary}
\end{table}
\endgroup

\section{Observations and Data Analysis}
\label{obs}
All the sources in our sample were observed as part of a joint {\it Fermi}-Gemini program (see footnote 1) using the 8.1\,$m$ Gemini-South telescope located in Chile between 06 February 2019 and 04 June 2019. The spectra were obtained using the Gemini Multi-Object Spectrographs (GMOS)\footnote{acquired through the Gemini Observatory Archive at NSF’s NOIRLab* and processed using the Gemini IRAF package} in the long-slit mode with the B600$-$G5323 grating (resolution, R$\sim$1700) and a slit width of 1.0 arcsec in the wavelength range 3500 \AA$-$7000 \AA.\\
All spectra reported here were obtained by combining at least three individual observations of the source with varying exposure times. This allowed us to reduce both instrumental effects and cosmic ray contribution. The data reduction was done following a standard procedure: the final spectra were all bias-subtracted, flat-normalized and corrected for bad pixels. 
The flat-field normalization is necessary to remove wavelength-dependent fluctuations that could affect the flat-field source spectrum.
After every observation of a source, we obtained a CuAr lamp spectrum to perform the wavelength calibration. This enables us to avoid potential shifts in the pixel-$\lambda$ calibration induced by the telescope motion during the night. Finally, all spectra were flux-calibrated using a spectroscopic standard, which were observed using the same 1.0$^{\prime\prime}$ slit used in the rest of the analysis. This data reduction and spectral extraction was done by using IRAF (PyRAF and Gemini IRAF, v1.14). 

\begin{figure*}
 \begin{minipage}[b]{.49\textwidth}
  \centering
  \includegraphics[width=0.9\textwidth]{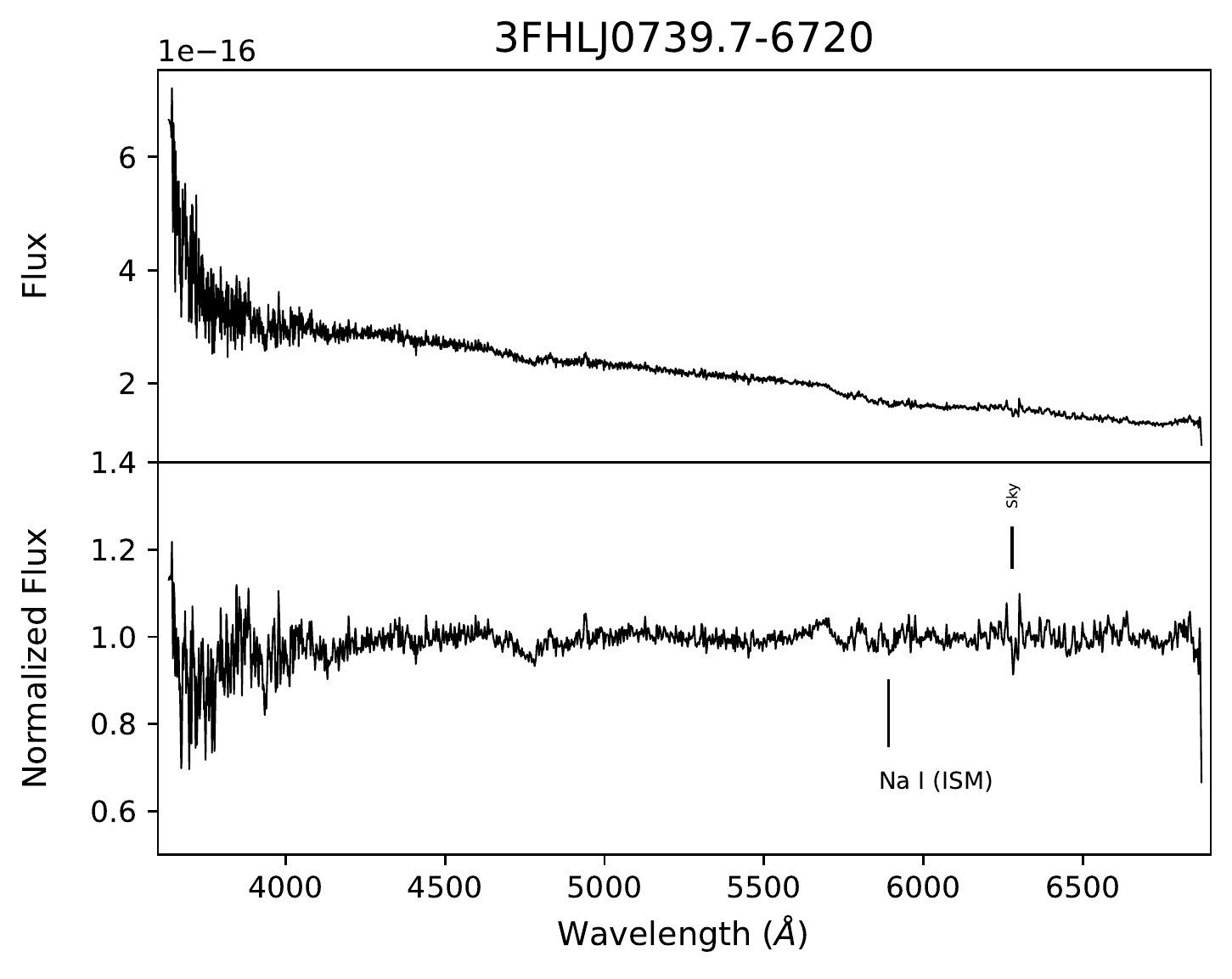}
  \end{minipage}
    \vspace{0.5cm}
\begin{minipage}[b]{.49\textwidth}
  \centering
  \includegraphics[width=0.9\textwidth]{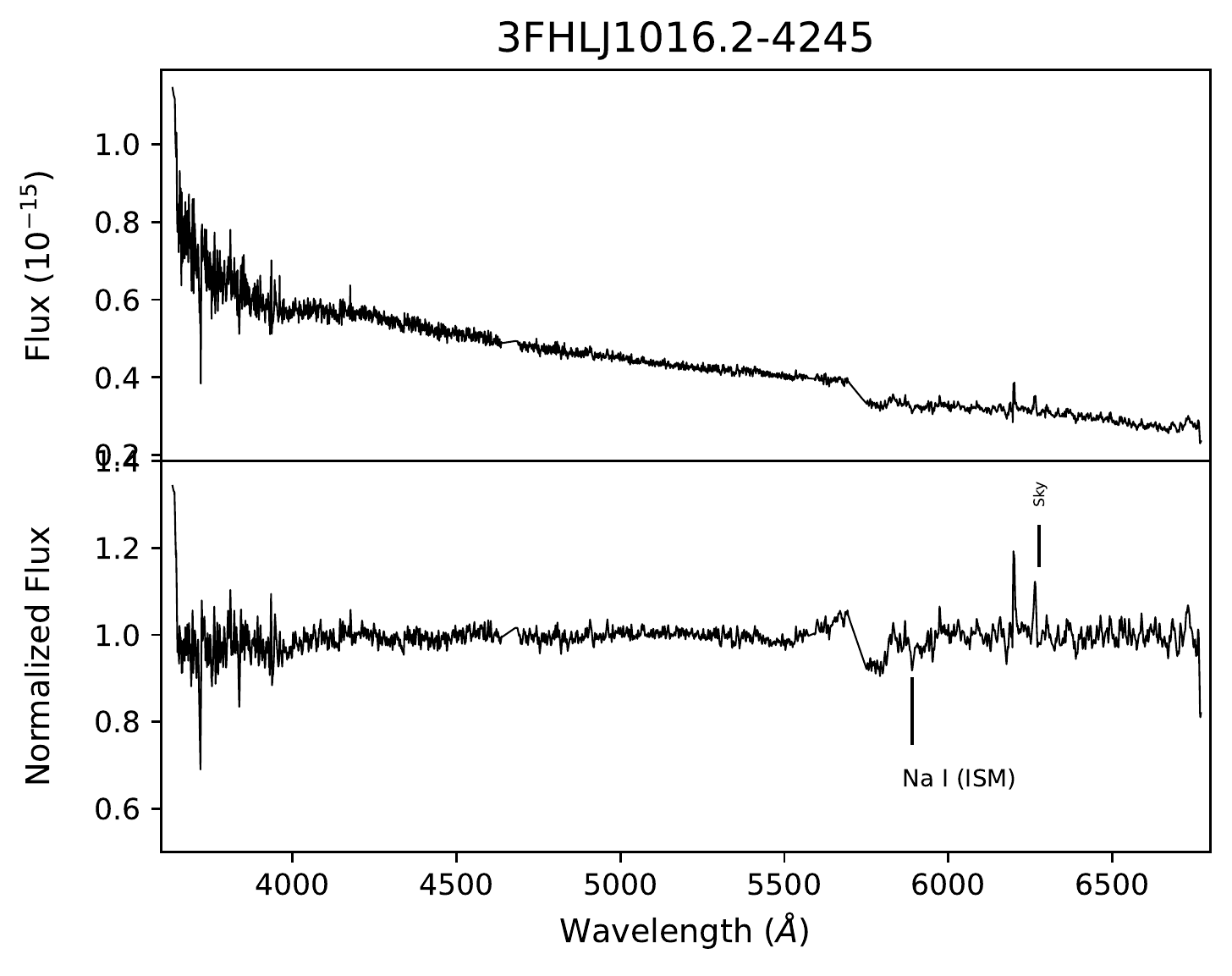}
  \end{minipage}
  \begin{minipage}[b]{.49\textwidth}
  \centering
  \includegraphics[width=0.9\textwidth]{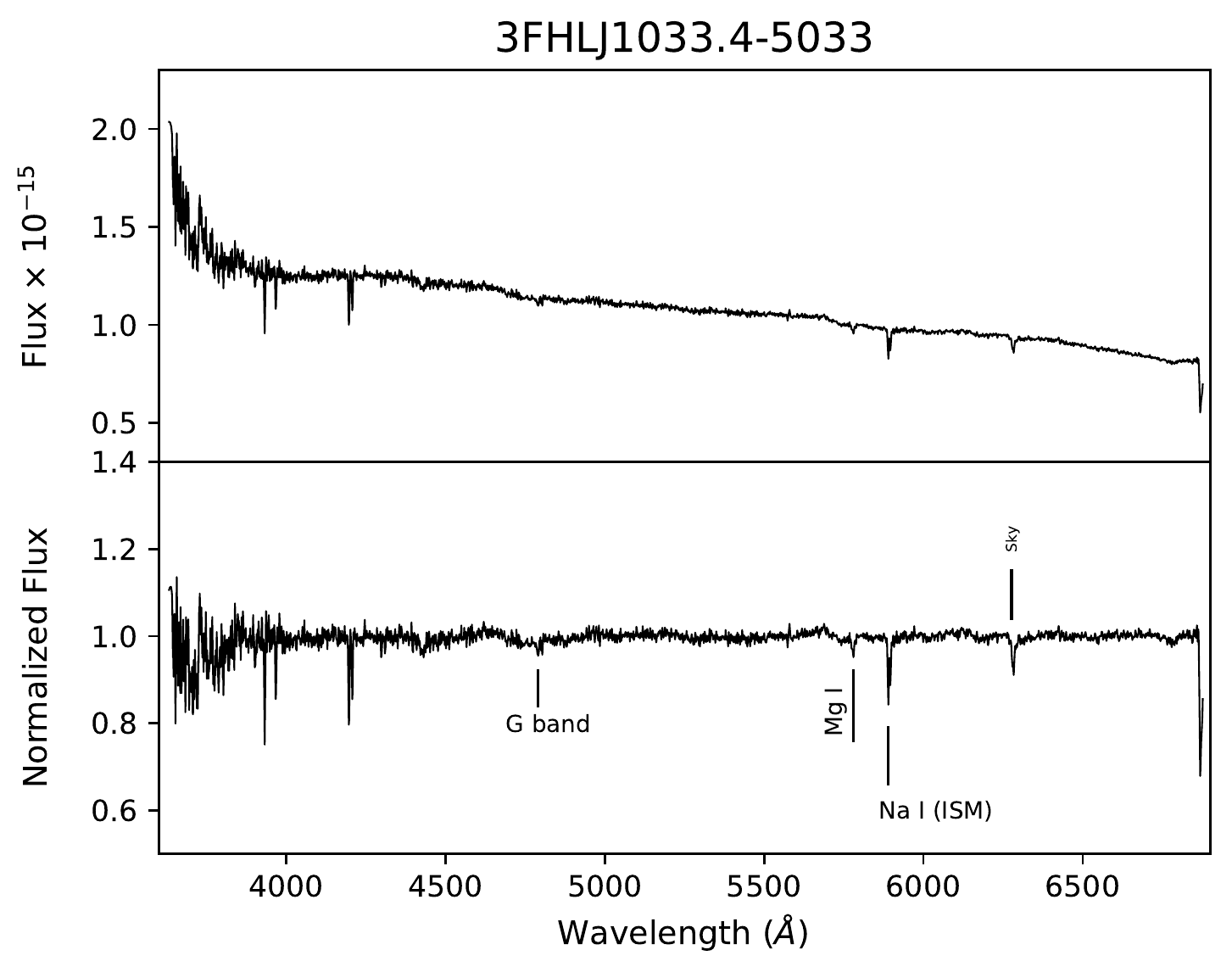}
  \end{minipage}
  \begin{minipage}[b]{.49\textwidth}
  \centering
  \includegraphics[width=0.9\textwidth]{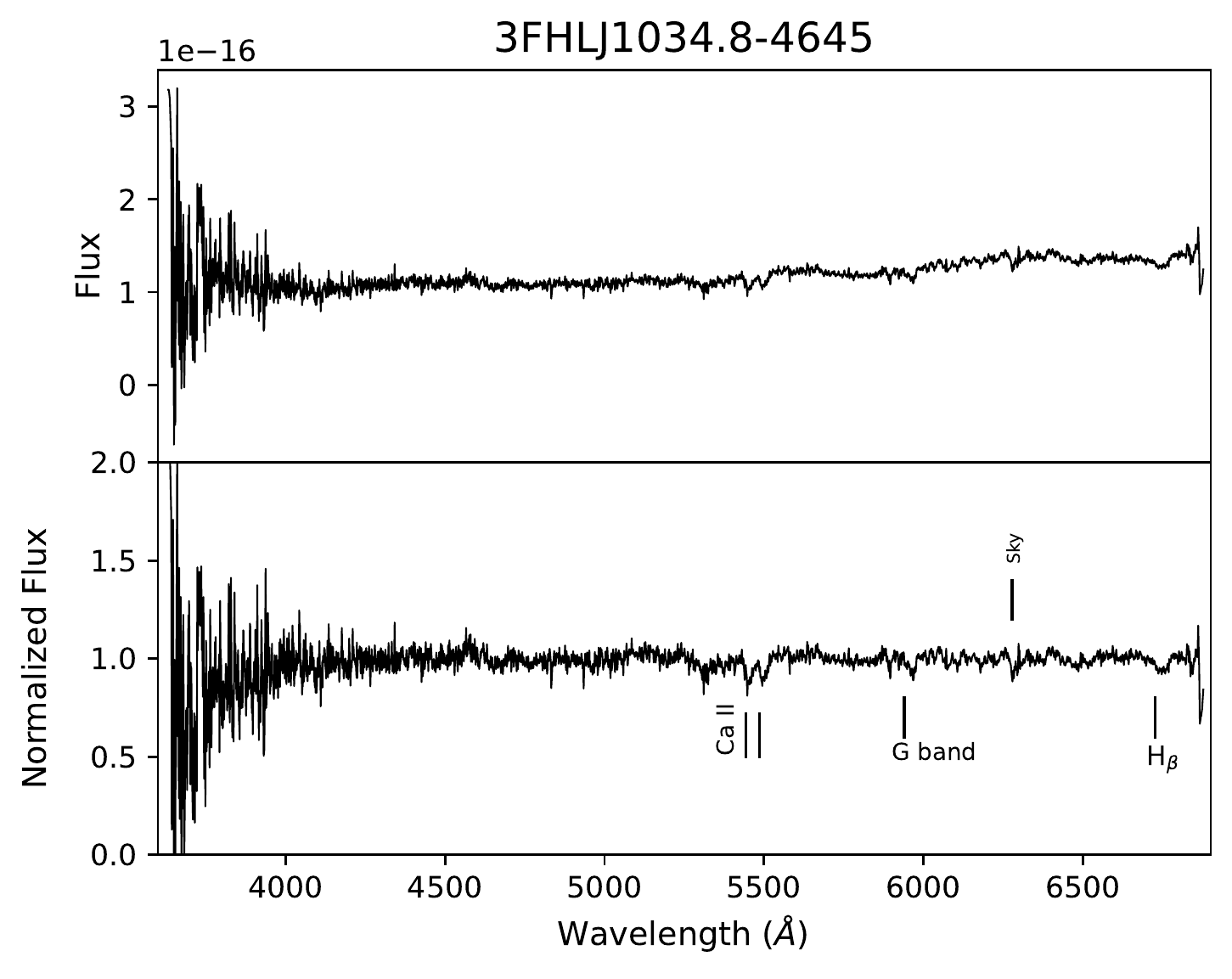}
  \end{minipage}
\begin{minipage}[b]{.49\textwidth}
  \centering
  \includegraphics[width=0.9\textwidth]{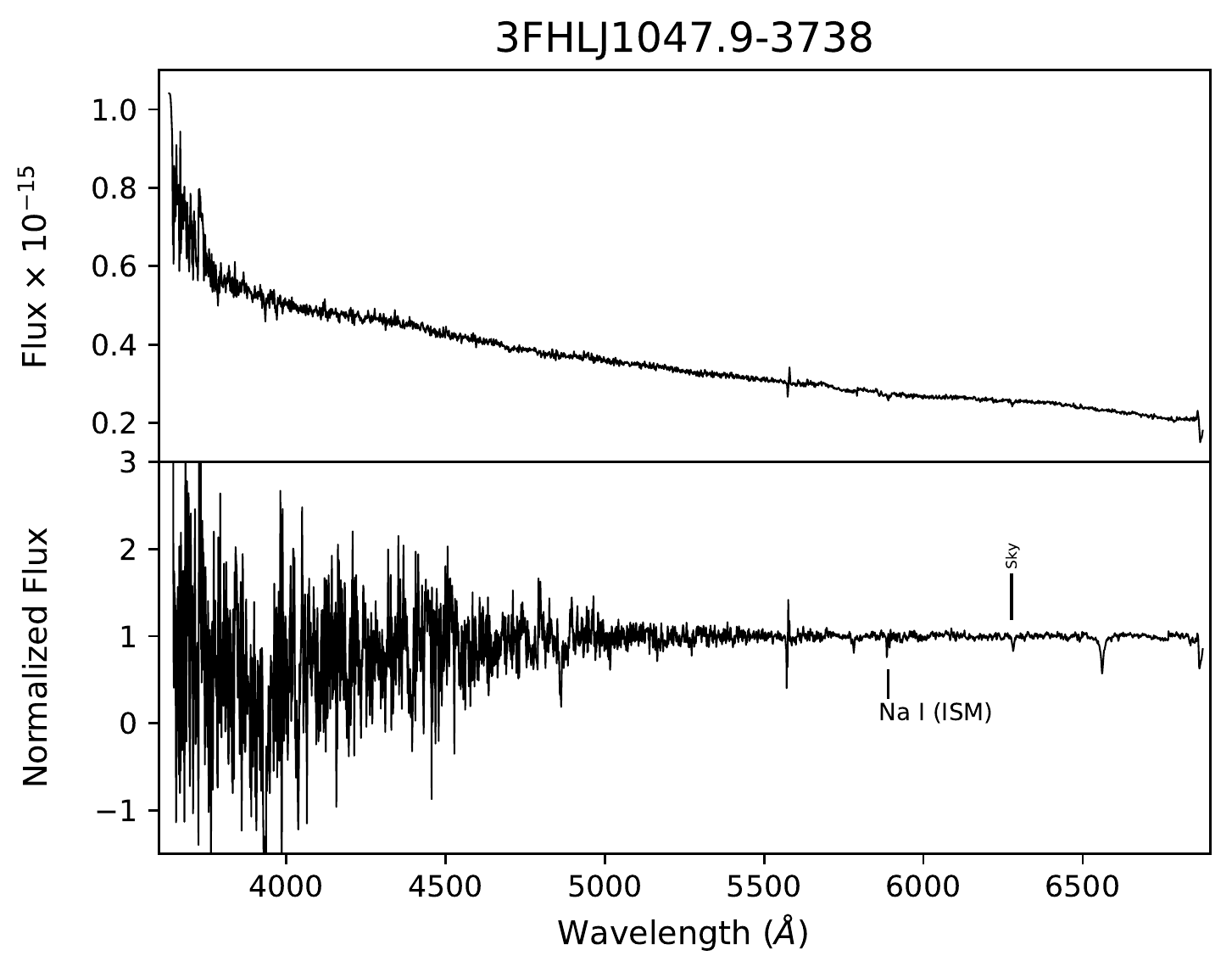}
  \end{minipage}
\caption{ Top panels: flux calibrated and de-reddened optical spectra of all our observed candidates. Bottom panels: normalized spectra. The absorption and/or emission features are labeled with the line element they represent. The $\otimes$ denotes the atmospheric features.}
\label{fig:spec}
\end{figure*}

\section{Spectral Analysis}
\label{spectral}

Table~\ref{tab:sample} contains the results from the spectral analysis (redshift and classification) of all analysed sources, and the corresponding spectra are shown in Figure~\ref{fig:spec}.

The continuum is taken to be a power-law unless the optical shape is more complex, in which case the preferred fit is described in Section~\ref{individual_src}. 
To make the absorption/emission features more apparent,  we also show the normalized spectra of our sources
obtained by dividing the flux-calibrated spectrum by the power-law fit of the continuum. The S/N of the normalized spectrum is then measured in a minimum of five individual featureless regions in the spectrum with a width of $\Delta\lambda\approx40$\,\r{A}. 

To find a redshift measurement, each spectrum was visually inspected for any absorption or emission feature. Any potential feature that matched known atmospheric lines\footnote{\url{https://www2.keck.hawaii.edu/inst/common/makeewww/Atmosphere/atmabs.txt}} was not taken into consideration. To test the reliability of any potential feature, its existence was verified in each of the individual spectral files used to obtain the final combined spectrum shown in Figure~\ref{fig:spec}.
Details for the sources are given in Section~\ref{individual_src}. The features described are also listed in Table~\ref{tab:sample} with the derived redshift measurement.

\subsection{Comments on Individual sources}\label{individual_src}
{\bf 3FHL J0739.7$-$6720:} This BCU is associated with the X-ray source 1RXS J073928.1$-$672147. No features are identified in the optical spectrum of the source, enabling us to classify it as a BL Lac object.\\

{\bf 3FHL J1016.2$-$4245:} This BCU is associated with the X-ray source 1RXS J101620.6$-$424733. The optical spectrum of this source was found to be featureless. This source is classified as a BL Lac object.\\

{\bf 3FHL J1033.4$-$5033:} This BCU is associated with the optical source 2MASS J10333216$-$5035287. In the optical spectrum of this source, the Na ISM line at 5890 \AA\, and another atmospheric feature at 6283 \AA\, are clearly visible. We were also able to identify Ca II (H \& K lines) at 3934 \AA\,, and 3969 \AA\,, respectively. This doublet is associated with the intervening medium since we observe them at rest wavelengths. Furthermore, we observe two absorption features at 5780 \AA\, and 4790 \AA\,, which can be attributed to Mg {\sc I} and G$-$band, respectively. This yields a redshift lower limit of $z>0.11$ and a BL Lac classification of the source.\\

{\bf 3FHL J1034.8$-$4645:} This BCU, associated with the source 1RXS J103438.7$-$464412, exhibited multiple absorption features in its optical spectrum: Ca {\sc II} (H \& K) lines at 5443.2 \AA\, and 5486.3 \AA, G$-$band at 5939.5 \AA, and H-$\beta$ at 6726 \AA. These features result in a redshift lower limit of $z> 0.38$ and a BL Lac classification of the source.\\

{\bf 3FHL J1047.9$-$3738:} This BCU is associated with the source 2WHSP J104756.8$-$373730. We observe a featureless optical spectrum for this source, thus, making it a BL Lac object.\\

\begingroup
\renewcommand*{\arraystretch}{1.8}
\begin{table*}\caption{Table of observed sources and their spectral properties. Column (1)-(2): 3FHL name \citep{ajello17} and associated counterpart (from radio/IR/optical/X-ray/radio surveys). Columns (3)-(4): right ascension (R.A., J2000) and declination (Dec., J2000). Column
(5): observation date. Column (6): exposure time (in seconds). Column (7): spectral signal to noise (S/N). Column (8): rest frame wavelength of the observed line. Column (9): observed wavelength of the same line. Column (10): line type (emission or absorption). Column (11): obtained redshift. Column (12): final source classification.}
\centering
\scalebox{0.65}{
\hspace{-3cm}
\begin{tabular}{lccccccccccc}
\hline
\hline
3FHL Name & Counterpart & R.A. & Dec & Obs Date & Exposure & S/N & Spectral Line & Observed $\lambda$ & Line type & Redshift & Classification\\
& & (hh:mm:ss) & (hh:mm:ss) & & (seconds) & & Rest frame $\lambda$ (\r{A}) & (\r{A}) & &\\
(1) & (2) & (3) & (4) & (5) & (6) & (7) & (8) & (9) & (10) & (11) & (12)\\
\hline
%3FHL J0002.1$-$6728 & SUMSS J000215$-$672653 & 00:02:15.21 & $-$67:26:52.91 & 0.0253 & 18.6 & June 1 2018 & 5400 & $-1.44$\\ 
3FHL J0739.7$-$6720 & 1RXS J073928.1$-$672147 & 07:39:27.39 & $-$67:21:36.4 & 24 Mar 2019 & 6366  & 80.16 & --- & --- & --- & --- & BL Lac\\
3FHL J1016.2$-$4245 & 1RXS J101620.6$-$424733 & 10:16:20.76 & $-$42:47:23.1 & 17 Mar 2019 & 3228  & 106.89 & --- & --- & --- & --- & BL Lac\\
3FHL J1033.4$-$5033 & 2MASS J10333216$-$5035287 & 10:33:32.11 & $-$50:35:27.1 & 06 Feb 2019 & 2836  & 157.60 & 4304 (G$-$band)& 4790 & Absorption & \\
& & & & & & & 5175 (Mg {\sc I}) & 5780 & Absorption & $> 0.11$ & BL Lac\\
3FHL J1034.8$-$4645 & 1RXS J103438.7$-$464412 & 10:34:38.49 & $-$46:44:03.5 & 24 Feb 2019 & 4032 & 56.87 & 3934 (Ca {\sc II}) & 5443.2 & Absorption & \\
&  & & & & & & 3969 (Ca {\sc II}) & 5486.3 & Absorption & \\
&  & & & & & & 4304 (G$-$band) & 5939.5 & Absorption & \\
&  & & & & & & 4861 (H$-\beta$) & 6726.7 & Absorption & $> 0.38$ & BL Lac\\
3FHL J1047.9$-$3738 & 2WHSP J104756.8$-$373730 & 10:47:56.94 & $-$37:37:30.8 & 06 Feb 2019 & 4032  & 121.45 & --- & --- & --- & --- & BL Lac\\
\hline 
\hline
\end{tabular}}
\label{tab:sample}
\end{table*}
\endgroup

\section{SED Modeling}\label{sedmodels}
We adopt a one-zone leptonic emission model to fit the SED of the two targets for which we could 
derive redshift constraints: 3FHL J1033.4$-$5033 ($z>0.11$) and 3FHL J1034.8$-$4645 ($z>0.38$).
The details of the full model can be found in \citet[][]{Ghisellini_2009}. In the following,
we provide 
few important guidelines. 

We assume that the entire SED is produced by a spherical region, located at $R_{\rm diss}$ from the central supermassive black hole, and that the
particles responsible for the emission are relativistic electrons. On the other hand, protons are assumed to be
cold, hence not radiating, and only contributing to the kinetic energy of the jet. Number densities of protons and electrons are assumed equal \citep[see][]{Celotti_2008} and contribution of pairs is not included in the model.
The electrons are distributed according to a broken power law: 
 \begin{equation}
 N(\gamma)  \, \propto \, { (\gamma_{\rm break})^{-p} \over
(\gamma/\gamma_{\rm break})^{p} + (\gamma/\gamma_{\rm break})^{q}}. \label{eq:shape}
\end{equation}

In the above, {\it p} and {\it q} are, respectively, the slopes before and after the energy break ($\gamma_{\rm break}$).
The entire region moves along the jet with a bulk Lorentz factor $\Gamma$ and is encompassed by a uniform magnetic field ($B$). 
The relativistic electrons are accelerated by the magnetic field and radiate via synchrotron process. This results in the non-thermal low-frequency SED
peak which extends from radio up to optical/X-rays. Further, if low-energy photons surrounds the emission region, these particles can 
undergo inverse Compton emission. When the photon field is the same produced by the synchrotron radiation, we refer to it as
synchrotron self Compton (SSC); when it is external to the jet (e.g. accretion disk, BLR, torus) we refer to it as external Compton (EC). 

The thermal components considered in the model are the accretion disk, the BLR and the torus. The accretion disk is modeled via a standard Shakura-Sunyaev disk \citep{shakura1973black} and its SED is explained by a multicolor black-body \citep{Frank_2002}. The BLR and the torus are 
modeled as spherical shells located at a distance $R_{\rm BLR} = 10^{17} L^{1/2}_{\rm disk,45}$ cm and $R_{\rm TORUS} = 10^{18} L^{1/2}_{\rm disk,45}$ cm (where $L_{\rm disk,45}$ is the disk luminosity in units of $10^{45}$\,erg s$^{-1}$). They are assumed to reprocess 10\% and 50\% of the disk emission, respectively, and their SED is modelled as a black-body peaking at the Lyman-$\alpha$ frequency and at 300\,K (typical torus temperature). The hot corona of electrons
above and below the accretion disk is modeled as power-law with exponential cut-off. It reprocess 30\% of the accretion disk radiation and upscatters photons up to $\sim200-500\,\rm keV$. 

The model computes the energy densities of all components, which depend on the distance of the emission region. The total jet power is computed as the sum of electron, proton, magnetic and radiative power. 

\subsection{The multi-band data}
Since the multi-wavelength data for our sources are scarce, to construct the SED we considered:
\begin{itemize}
\item At $\gamma$-rays: data from the 3FHL \citep{ajello17} and the 4FGL-DR3 \citep{4FGL_2022}.
\item At X-rays: both sources have archival {\it Swift}-XRT observations. We analysed them with the standard \texttt{xrtpipeline} and extracted their source and background spectra via \texttt{xselect}. For both, we considered a 20\arcsec~circle for the source region and an annulus of 50\arcsec~and 100\arcsec~inner and outer radius centered on the source position as the background. The ancillary response files were created with \texttt{xrtmkarf} and the spectra were rebinned with \texttt{grppha} considering 10 counts per bin. An absorbed power-law model (\texttt{tbabs*po}) with absorption fixed to its Galactic value \citep{Kaberla_2005} was then used in XSPEC to fit the X-ray spectra and extract the X-ray photon index and flux. Table~\ref{tab:x-ray_spec} lists the derived X-ray spectral parameters.
\item At UV: as {\it Swift}-UVOT observe contemporaneously with XRT, we checked if the sources were detected by any of the UVOT filters using the \texttt{uvotdetect} task. Only 3FHL 1034.8-4645 was detected by the \texttt{uvw1} filter (see Table~\ref{tab:x-ray_spec}). 
\item Archival observation (obtained using the SED Builder\footnote{\url{ https://tools.ssdc.asi.it/SED/}}) allowed us to collect further multiband data and complete the SED from radio up to $\gamma$-rays. 
\end{itemize}

\begingroup
\renewcommand*{\arraystretch}{1.5}
\begin{table}[t!]
\centering
\
\caption{Table of Swift observations and derived spectral parameters.}
\scalebox{0.9}{
\hspace{-2cm}
\begin{tabular}{c|ccccccc}
\hline
& obs ID & obs date & obs length & $N_{\rm H}$ & $\Gamma_{0.3-10\,\rm keV}$ & $\rm Flux_{\rm unabs, 0.3-10\, keV}$ & $\rm Flux_{\texttt{uw1}}$\\
& & & [ks] & $[\rm cm^{-2}]$ &  & $[\rm erg~cm^{-1}~s^{-1}]$ & $[\rm erg~cm^{-1}~s^{-1}]$  \\
\hline
3FHL J1033.4$−$5033 & 00041356001 & 29 Sep 2010 &  3.90 & $1.70\times10^{21}$ & $2.08\pm0.55$ & $1.85_{-0.19}^{+0.44}\times10^{-12}$ & --- \\   
\hline
3FHL J1034.8$−$4645 & 00046768001 & 13 Jan 2012 & 1.75 & $1.35\times10^{21}$ & $1.95\pm0.61$ & $3.87_{-0.65}^{+0.91}\times10^{-12}$ & $(1.46\pm0.08)\times10^{-12}$ \\
\hline
\end{tabular}}
\label{tab:x-ray_spec}
\end{table}
\endgroup

\subsection{Constraints to the model}

Both our sources show no evidence of emission lines in their optical spectra (see Figure\,\ref{fig:spec} and Section\,\ref{spectral}). Therefore they are canonically 
classifiable as BL Lacs. The SED of such blazars is usually explained by the SSC scenario, possibly owing to a low-power accretion disk which 
is not capable of providing substantial radiation for the EC (see \citealp{Giommi2002,Padovani2002} for an alternative view). The sources peak positions, their redshifts, and derived $\gamma$-ray and X-ray luminosities also point to a low-power jet, resembling BL Lacs sources. We therefore test only a model with synchrotron and SSC emission processes. 

As their optical spectra are featureless, constraints on the disk luminosities and black hole masses are hard to obtain. Therefore, we rely on empirical relations \citep{sbarrato2012,ghisellini2012blue} between the  $\gamma$-ray and BLR luminosity: $L_{\rm BLR}\sim4L_{\gamma}^{0.93}$.
For 3FHL J1033.4$-$5033, assuming the lower limit of $z>0.11$ as our fiducial redshift, $L_{\gamma}\sim8\times10^{43}$\,erg s$^{-1}$, hence $L_{\rm BLR}\sim2.7\times10^{41}$\,erg s$^{-1}$; for  3FHL J1034.8$-$4645 ($z>0.38$), $L_{\gamma}\sim3\times10^{44}$\,erg s$^{-1}$, hence $L_{\rm BLR}\sim9\times10^{41}$\,erg s$^{-1}$.
Under the assumption that the BLR reprocesses 10\% of the disk emission, this implies $L_{\rm disk}\sim2\times10^{42}$\,erg s$^{-1}$ and  $L_{\rm disk}\sim9\times10^{42}$\,erg s$^{-1}$, respectively.
As for the black-hole mass, we employ an average of $<M_{\rm BH}>=10^8\,\rm M_{\odot}$.

\begin{figure} 
\centering
    \includegraphics[width=0.45\textwidth]{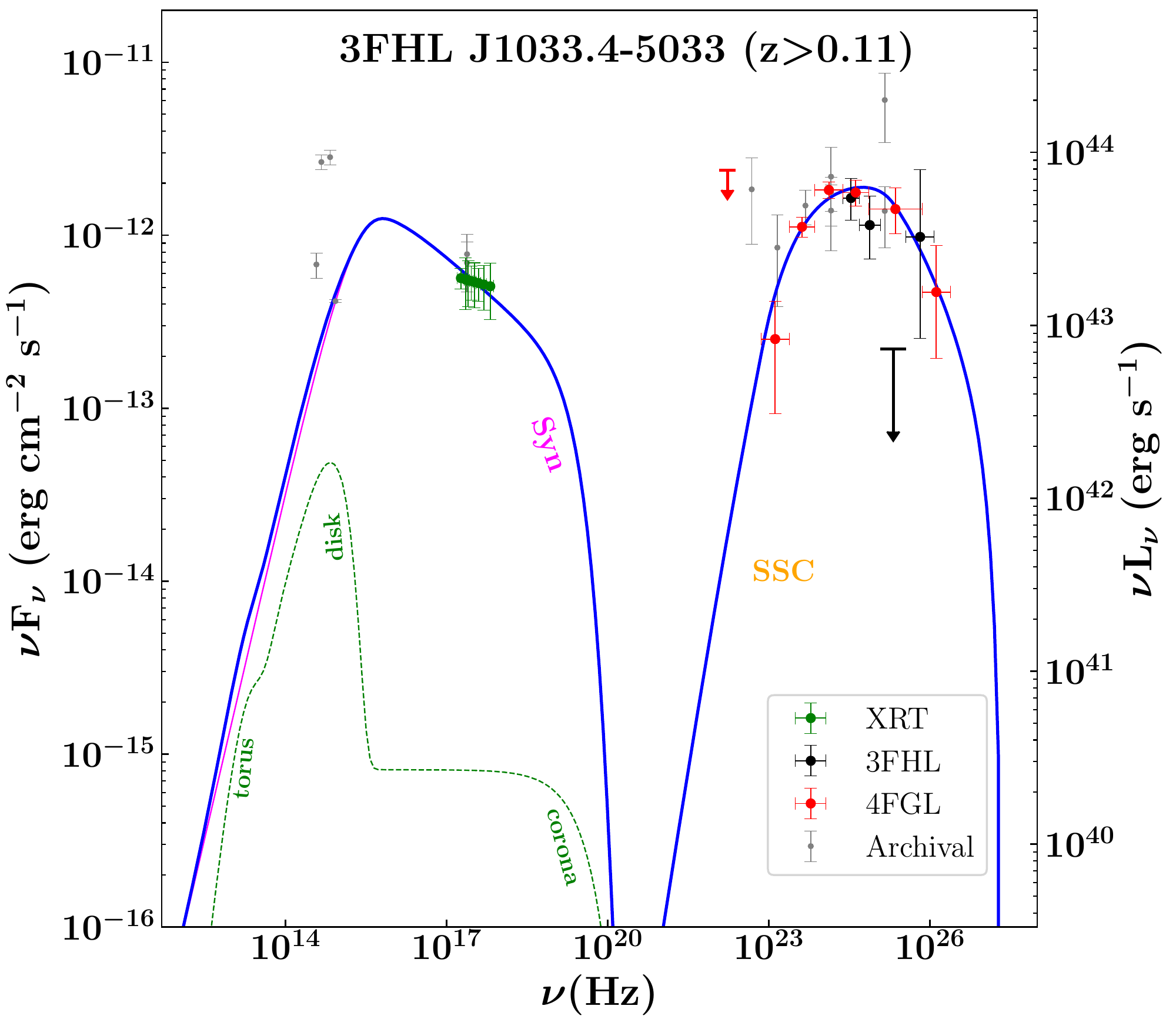}
    \includegraphics[width=0.45\textwidth]{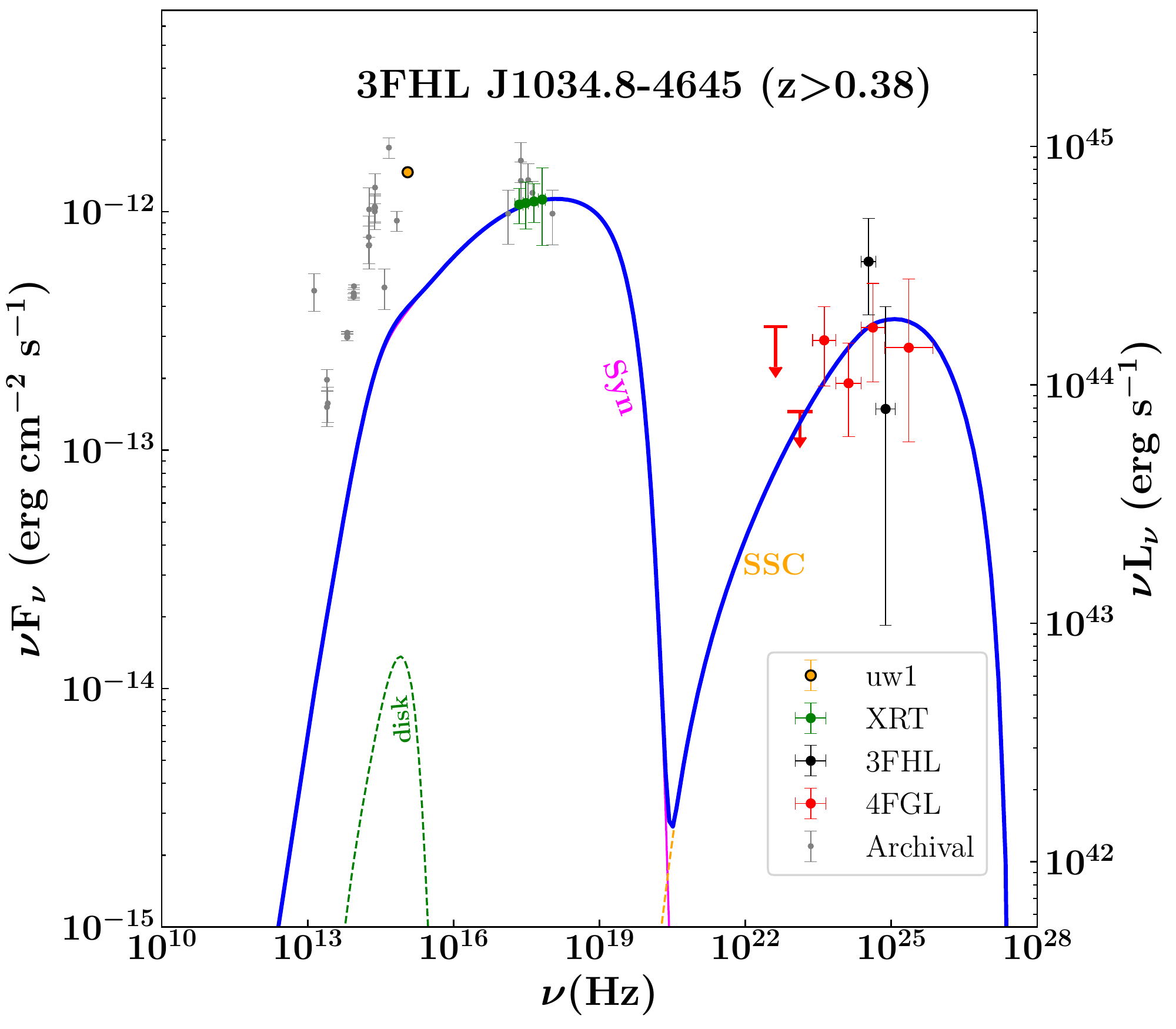}
    \includegraphics[width=0.45\textwidth]{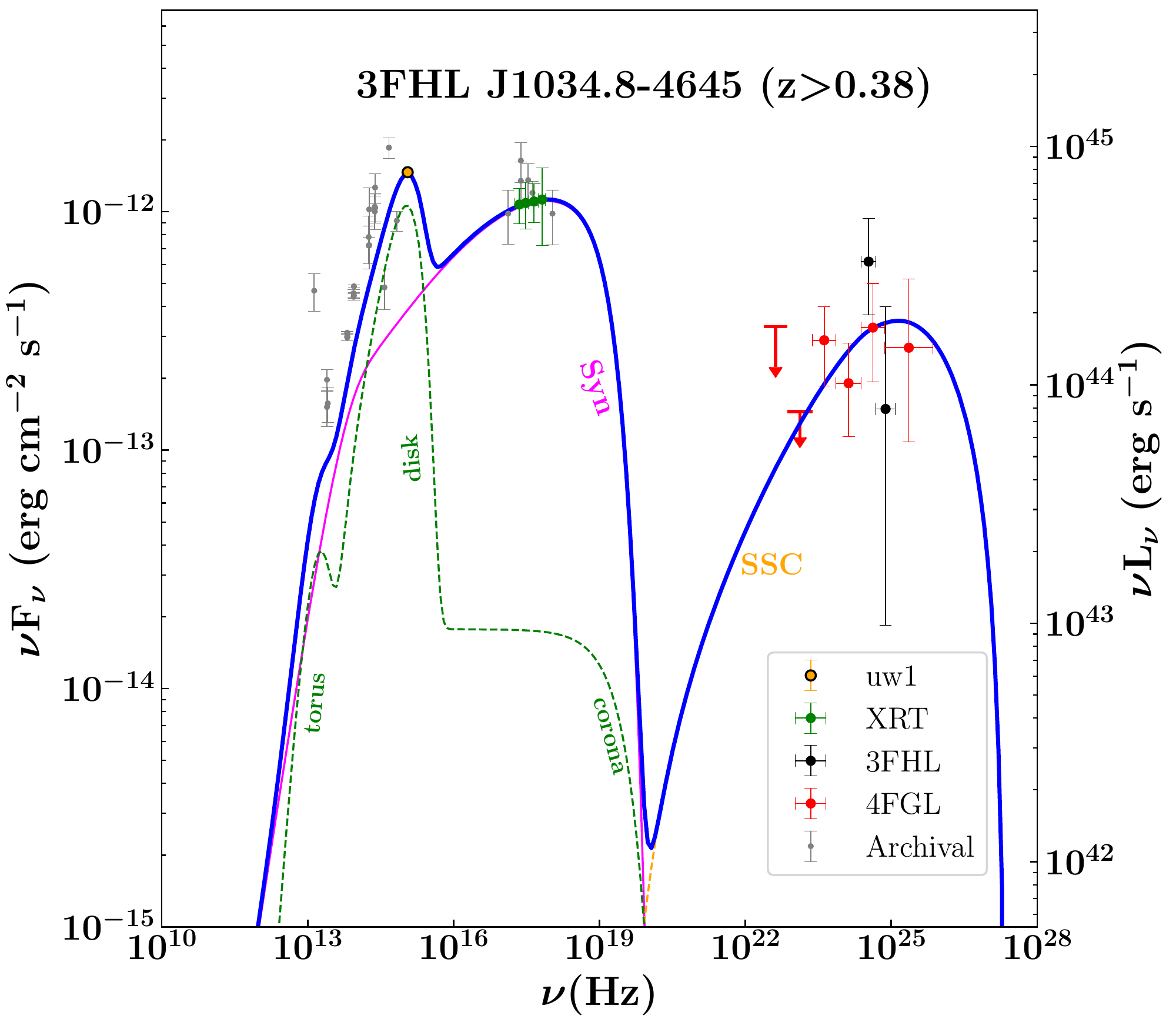}
    \includegraphics[width=0.45\textwidth]{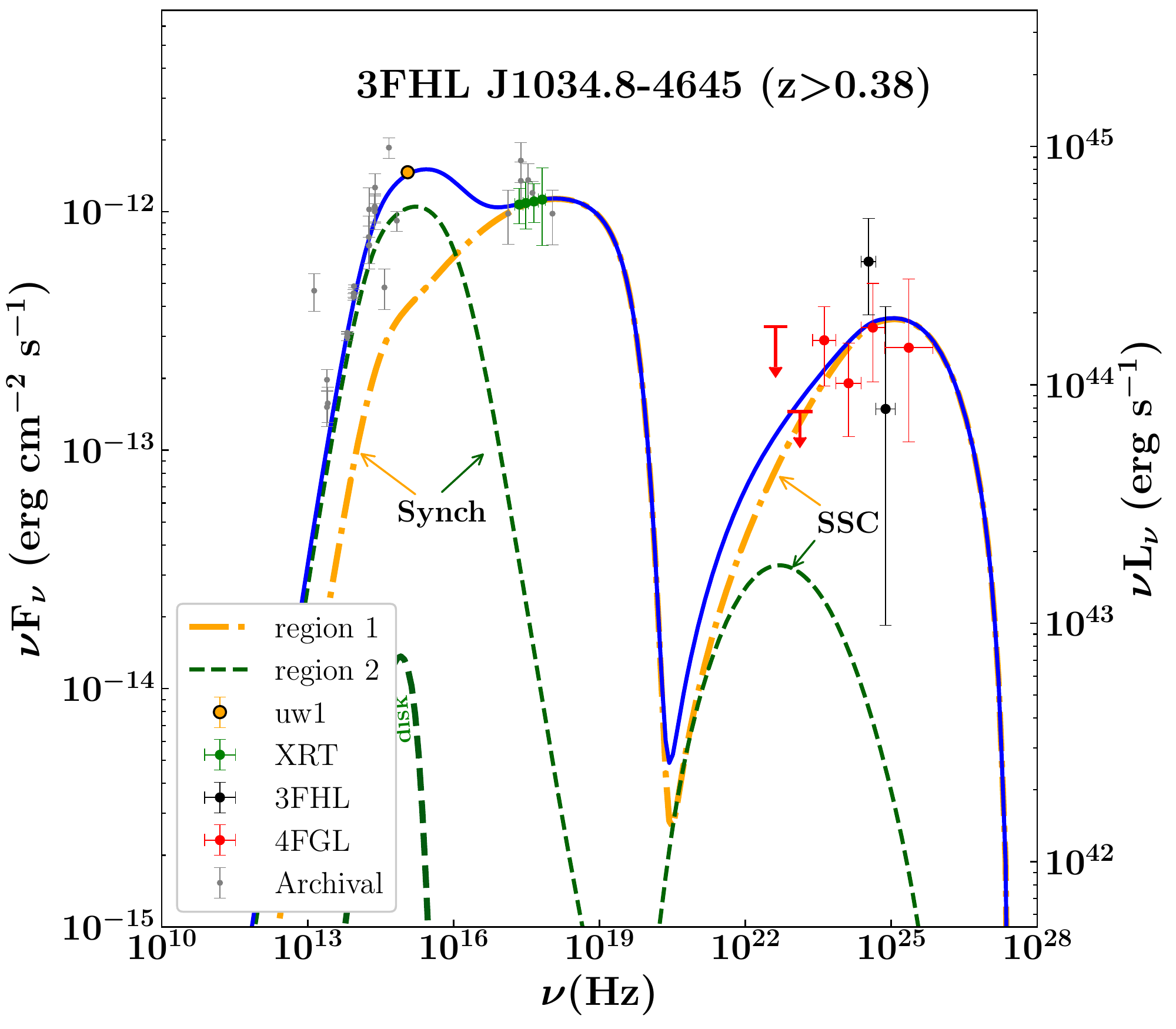}
        \caption{{\it Top panel:} SED of 3FHL J1033.4$-$5033 ($z>0.11$, left) and 3FHL J1034.8$-$4645 ($z>0.38$, right). The data points are the multi-band data collected for the sources (from radio up to $\gamma$-rays). The lines are the various model components, with the synchrotron emission labelled in pink, the SSC in yellow and the thermal components in green. The total SED model is shown by the solid blue line. {\it Bottom panel}: SED of 3FHL J1034.8-4645 ($z>0.38$) with higher black hole mass and accretion disk luminosity (left); two emission regions (right).
         \label{fig:sed_3fhl_j1033}}
\end{figure} 

\subsection{Modeling results}
The top panels of Figure\,\ref{fig:sed_3fhl_j1033} show the broad-band SED for the 2 targets and Table~\ref{tab:sed_par} reports the best-fit parameter values.
From a modeling perspective, both sources resemble typical BL Lac objects. Their radiative power is dominant with respect to the kinetic and magnetic ones and the Doppler factors are within $8-11$ and the magnetic fields are $\sim1.1-2.2$ Gauss. The location region is within the BLR region for 3FHL J1033.4-5033, while slightly further for 3FHL J1034.8-4645. We note that for both sources we only have a lower limit on the redshift. Nonetheless, these sources are unlikely to be located away in time as the absorption from the extragalactic background light (EBL, \citealp{EBL_2018}) would start appearing in the $\gamma$-ray spectrum of these sources as a sharp attenuation at $E>10\,\rm GeV$. This is not seen in either $\gamma$-ray spectra.

\begin{table}[t!]
\begin{center}
\caption{Summary of the parameters used/derived for the one-zone leptonic SED modeling of 3FHL J1033.4$-$5033 and 3FHL J1034.8$-$4645. A viewing angle of 3$^{\circ}$ and 1$^{\circ}$ are adopted, respectively.}\label{tab:sed_par}
\hspace{-1.5cm}
%\resizebox{\textwidth}{!}{
\begin{tabular}{lcc}
\hline
Parameter & 3FHL J1033.4$-$5033 & 3FHL J1034.8$-$4645 \\
\tableline
\tableline
Synchrotron Peak frequency [Hz]   &  $8\times10^{14}$ & $2.5\times10^{17}$\\
Black hole mass ($M_{\rm BH}$) in log scale [$\rm M_{\odot}$]    & 8 & 8\\
Accretion disk luminosity ($L_{\rm disk}$) in log scale [erg s$^{-1}$]   & 42.30 & 42.95\\
Accretion disk luminosity in Eddington units ($L_{\rm disk}/L_{\rm Edd}$)   & $10^{-4}$ & $7\times10^{-4}$\\
Size of the BLR ($R_{\rm BLR}$) [pc ($R_{\rm Sch}$)] & $1.5\times 10^{-3}$ (151.43) & $3.0\times 10^{-3}$ (321.24)  \\
Dissipation distance ($R_{\rm diss}$) [pc ($R_{\rm Sch}$)] & $1.91\times10^{-3}$ (194.03) & $3.3\times10^{-3}$ (349.99)\\
Slope of the particle distribution below the break energy ($p$)   & 1.50 & 2.40 \\
Slope of the particle distribution above the break energy ($q$)   & 3.55 & 3.30  \\
Magnetic field ($B$) [G]                                          & 1.1  & 2.2 \\
Particle energy density ($U_{e}$) [erg cm$^{-3}$]                 & 0.27 & 0.06 \\
Bulk Lorentz factor ($\Gamma$)                                    & 8.16 & 10.80 \\
Minimum Lorentz factor ($\gamma_{\rm min}$)                       & $5\times10^3$ & $10^3$  \\
Break Lorentz factor ($\gamma_{\rm break}$)                       & $4.5\times{10^4}$ & $5.1\times{10^4}$ \\
Maximum Lorentz factor ($\gamma_{\rm max}$)                       & $5\times10^5$ & $5\times10^5$ \\ %\hdashline
\hline
\hline
Jet power in electrons ($P_{\rm e}$) in log scale  [erg s$^{-1}$]      & 41.77 & 41.86 \\
Jet power in magnetic field ($P_{\rm B}$) in log scale [erg s$^{-1}$]  & 41.01 & 42.35\\
Radiative jet power ($P_{\rm r}$) in log scale [erg s$^{-1}$]          & 42.79 & 43.38\\
Jet power in protons ($P_{\rm p}$) in log scale [erg s$^{-1}$]         & 41.09 & 41.69\\
Total jet power ($P_{\rm TOT}$) in log scale [erg s$^{-1}$]            & 42.84 & 43.44\\
\hline
\end{tabular}
%}
\end{center}
\end{table}

\begin{table}[t!]
\begin{center}
\caption{Summary of the parameters used/derived for the SED modeling 3FHL J1034.8$-$4645 ($z>0.38$) to explain the \texttt{uw1} data point. We consider: (1) higher black hole mass and accretion disk luminosity; (2) two emission regions in the jet. The parameters of the first region are the same as listed in Table~\ref{tab:sed_par}. A viewing angle of 1$^{\circ}$ is adopted, respectively.}\label{tab:sed_par_extra}
\hspace{-1.5cm}
%\resizebox{\textwidth}{!}{
\begin{tabular}{l|c|cc}
\hline
Parameter & Disk & Region 2  \\
\tableline
\tableline
Synchrotron Peak frequency [Hz]   & $2.5\times10^{17}$  & $4\times10^{15}$\\
Black hole mass ($M_{\rm BH}$) in log scale [$\rm M_{\odot}$]    & 8.69 & 8\\
Accretion disk luminosity ($L_{\rm disk}$) in log scale [erg s$^{-1}$]   & 44.84 & 42.95\\
Accretion disk luminosity in Eddington units ($L_{\rm disk}/L_{\rm Edd}$)   & $10^{-2}$ & $7\times10^{-4}$\\
Size of the BLR ($R_{\rm BLR}$) [pc ($R_{\rm Sch}$)] & $2.69\times 10^{-2}$ (566.63) & $3.0\times 10^{-3}$ (321.24)  \\
Dissipation distance ($R_{\rm diss}$) [pc ($R_{\rm Sch}$)] & $1.19\times10^{-2}$ (249.90) & $1.1\times10^{-2}$ (1200)\\
Slope of the particle distribution below the break energy ($p$)   & 2.40 & 2.40 \\
Slope of the particle distribution above the break energy ($q$)   & 3.30 & 5.50  \\
Magnetic field ($B$) [G]                                          & 0.8  & 1.2 \\
Particle energy density ($U_{e}$) [erg cm$^{-3}$]                 & 0.01 & 0.003 \\
Bulk Lorentz factor ($\Gamma$)                                    & 9.12 & 15.0 \\
Minimum Lorentz factor ($\gamma_{\rm min}$)                       & $1\times10^3$ & $10^3$  \\
Break Lorentz factor ($\gamma_{\rm break}$)                       & $9.3\times{10^4}$ & $7.6\times{10^3}$ \\
Maximum Lorentz factor ($\gamma_{\rm max}$)                       & $5\times10^5$ & $5\times10^5$ \\ %\hdashline
\hline
\hline
Jet power in electrons ($P_{\rm e}$) in log scale  [erg s$^{-1}$]      & 42.19 & 41.95 \\
Jet power in magnetic field ($P_{\rm B}$) in log scale [erg s$^{-1}$]  & 42.43 & 43.18\\
Radiative jet power ($P_{\rm r}$) in log scale [erg s$^{-1}$]          & 43.47 & 42.78\\
Jet power in protons ($P_{\rm p}$) in log scale [erg s$^{-1}$]         & 42.01 & 41.89\\
Total jet power ($P_{\rm TOT}$) in log scale [erg s$^{-1}$]            & 43.54 & 43.35\\
\hline
\end{tabular}
%}
\end{center}
\end{table}

The good spectral coverage at $\gamma$-rays for 3FHL J1033.4-5033 enables us to constrain the location of the high-energy peak position, which in turn locks the peak location of the synchrotron component at $\sim8\times10^{14}\,\rm Hz$. Moreover, the $\gamma$-ray spectrum allows us to put constraints on the shape of the particle population and on their Lorentz factor distributions ($\gamma_{\rm min}$ and $\gamma_{\rm max}$). Owing to the large uncertainties on the X-ray spectrum, we favor constraints from the $\gamma$-rays to fit the SED. Within the statistical uncertainties it can be seen that the model explains well the XRT spectrum. 

For 3FHL J1034.8-4645 we face the reverse challenge. While the $\gamma$-ray spectra suffers from large uncertainties, the X-ray spectrum is well constrained. We therefore rely on the low energy data to constrain the SED parameters. The X-ray spectral index is quite flat ($\Gamma_{\rm X}\sim1.9$), sampling the SED close to the peak of the synchrotron emission, derived to be at $\sim2.5\times10^{17}\,\rm Hz$. The flatness of the X-ray spectrum also implies a very flat index for the low energy particle population ($p=2.4$). Interesting for this particular source is the UVOT data point, which is strictly contemporaneous with the XRT observation. Blazars are known to be variable sources and therefore one needs to be cautious while using archival data to constrain the physical parameters. In this case, the UVOT point agrees well with the level of emission from archival observation, and it is a factor of 2 higher in flux than the X-ray detection. Therefore we tried to use our physical model to explore two scenarios: (1) the UVOT and archival data are sampling the disk emission; (2) the UVOT and archival points are produced by a second region located further along the jet. For the first case, we raise the level of the disk emission until it matched the UVOT point and increased the black hole mass to match the peak position. For the second case, we consider region 1 to be the same as the one derived in the one-zone case (Table~\ref{tab:sed_par}); region 2 instead is considered to be located further along the jet and modeled with the physical SED described above. The sum of the 2 regions give us the total SED. Both scenarios are shown in the bottom panels of  Figure~\ref{fig:sed_3fhl_j1033} and the corresponding model parameters for the disk case and the second region are reported in \ref{tab:sed_par_extra}. 

It can be seen that both scenarios explain fairly well the total SED. The accretion disk case is challenging to support, though it is a very interesting possibility as it would be the first detection of an accretion disk emission from a BL Lac source. Indeed, if this was the case, the UVOT data would be produced by a strong accretion disk which is not swamped by the non-thermal emission. According to the unification scenario \citep{urry95}, we would therefore expect to see some broad lines in the optical spectrum. Absence of such lines could point to either the absence of the broad line region clouds, or to the fact that, though stronger, the accretion disk emission is not sufficient to ionize the clouds. The second scenario could be more plausible as AGN jets detected at radio frequencies are known to be knotted \citep[e.g.][]{King_2016}. According to our best-fit, the two regions would be separated by $\sim850\,\rm R_{\rm sch}$, both still within the torus. The synchrotron peak of the second region should be at $\sim4\times10^{15}\,\rm Hz$ and the SSC component would be subdominant in the high-energy regime. The magnetic field of the second region would be lower ($B\sim1.2$) and the $\Gamma$ factor higher ($\Gamma=15$). With regards to the jet powers, we find that both scenarios would still fit the BL Lac parameter space and the combined total jet power is very similar in both cases.

A third plausible scenario to explain the excess emission at the lowest frequencies is thst we could be seeing the emission from the host galaxy \citep[e.g.,][]{ghisellini17, Archer_2018, Rosillo_2022}. Fitting the IR to UV data with a single temperature blackbody, we obtain a $T_{\rm eff, host}\sim10^4\,\rm K$ and a luminosity of $L_{\rm host}\sim 10^{45}\,\rm erg~cm^{-2}~s^{1}$. Future multi-band follow-up observations of these source will be crucial to disentangle these scenarios.

\begin{figure*} 
\centering
        \includegraphics[width=0.49\columnwidth]{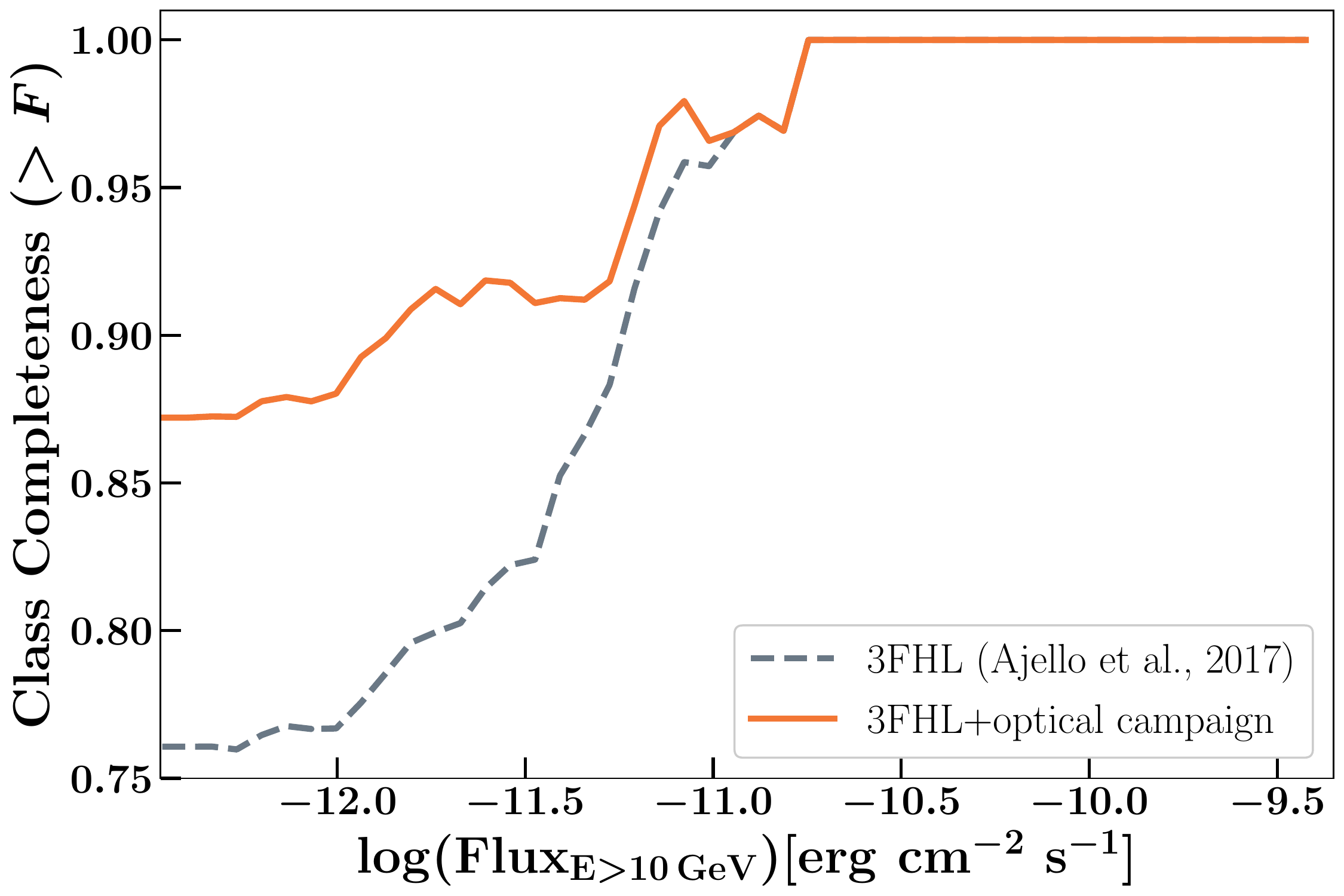}
        \includegraphics[width=0.49\columnwidth]{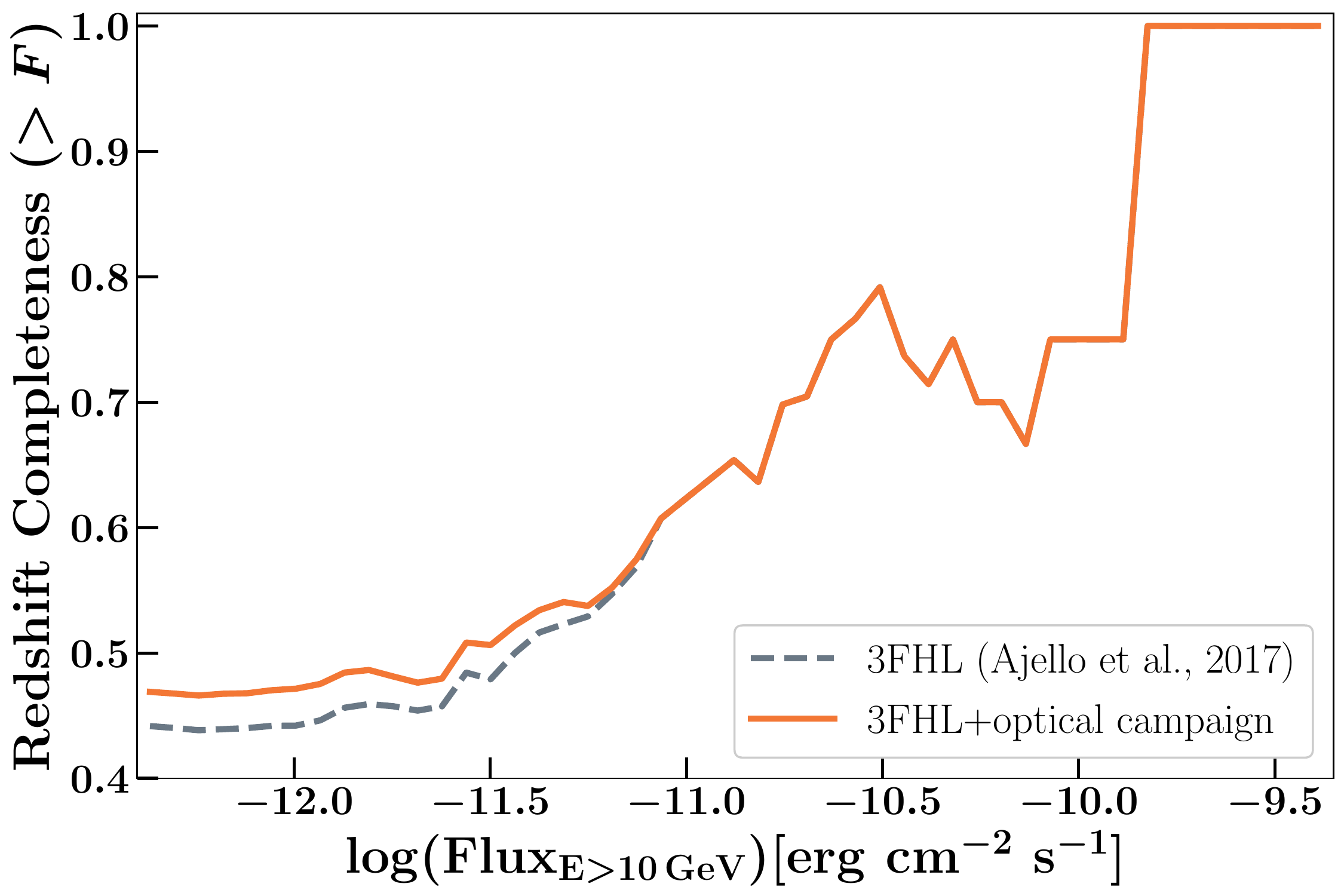}
        \caption{{\bf Left}: classification completeness as function of energy flux for all the identified blazars in the 3FHL. The gray dotted line represents the completeness of the original catalog, while the orange solid line represents the completeness after the optical campaign efforts undertaken by our group. As can be seen, the fraction of unclassified 3FHL blazars has been reduced from 23\% to 11\%.
         {\bf Right}: redshift completeness as function of energy flux for all the identified blazars in the 3FHL. The improvement obtained thanks to the optical spectroscopic follow-up is of $\sim 3\%$ (orange solid line) with respect to the original 3FHL.
         \label{fig:completeness}}
\end{figure*}

\section{Conclusion}

{ In this work, we present the results the optical spectroscopic campaign directed towards  rendering the 3FHL blazar sample spectroscopically complete using the Gemini Multi-Object Spectrographs (GMOS) mounted on the 8.1\,$m$ Gemini-South telescope in Chile.} We observed 5 extragalactic sources classified as BCU (blazars of uncertain classification) in the 3FHL catalog.
All the objects in our source sample are classified as BL Lacs based on their observed optical spectrum. 
Out of the 5 BCUs in our sample we found redshift lower limits for two source, and featureless spectra with no {redshift} measurement for the remaining 3 sources. Having obtained redshift lower limits for two of our sources, we were also able to collect their multi-band data and test a one-zone leptonic model.  The results show that both sources well agree with being standard BL Lac objects. 

In combination with the other campaign results, we were able to decrease (i) the fraction of blazars without a classification from  $\sim$23\% to $\sim$11\%, and (ii) blazars without redshift from $\sim$56\% to $\sim$53\%. These results are highlighted in Figure~\ref{fig:completeness}, where on the right we show the significant improvement in classification completeness of the 3FHL catalog as function of energy flux, and to the right the increase on redshift completeness. Table~\ref{tab:summary} summarizes the major results from this campaign. Future progress on making the 3FHL blazar sample  spectroscopically complete will rely on granted telescope time by $4-8\,m$ facilities and/or ongoing spectroscopical follow-ups aiming at completing complementary catalogs (such as the 3FGL or the 4FGL, see \citealp{Pena-Herazo_2021}).

\label{conclusion}

\section{Acknowledgements}

This work relied on the use of the TOPCAT software \citep{taylor05} for the analysis of data tables.\\
 Based on observations obtained at the international Gemini Observatory, a program of NSF’s NOIRLab, which is managed by the Association of Universities for Research in Astronomy (AURA) under a cooperative agreement with the National Science Foundation. on behalf of the Gemini Observatory partnership: the National Science Foundation (United States), National Research Council (Canada), Agencia Nacional de Investigaci\'{o}n y Desarrollo (Chile), Ministerio de Ciencia, Tecnolog\'{i}a e Innovaci\'{o}n (Argentina), Minist\'{e}rio da Ci\^{e}ncia, Tecnologia, Inova\c{c}\~{o}es e Comunica\c{c}\~{o}es (Brazil), and Korea Astronomy and Space Science Institute (Republic of Korea).
SM acknowledges funding from the INAF ``Progetti di Ricerca di Rilevante Interesse Nazionale'' (PRIN), Bando 2019 (project: ``Piercing through the clouds: a multiwavelength study of obscured accretion in nearby supermassive black holes''). 
L.M acknowledges that support for this work was provided by NASA through the NASA Hubble Fellowship grant \#HST-HF2-51486.001-A, 
awarded by the Space Telescope Science Institute, which is operated by the Association of Universities 
for Research in Astronomy, Inc., for NASA, under contract NAS5-26555.

\bibliographystyle{aasjournal}
\bibliography{3FHL_ctio}

\begin{thebibliography}{}
\expandafter\ifx\csname natexlab\endcsname\relax\def\natexlab#1{#1}\fi
\providecommand{\url}[1]{\href{#1}{#1}}
\providecommand{\dodoi}[1]{doi:~\href{http://doi.org/#1}{\nolinkurl{#1}}}
\providecommand{\doeprint}[1]{\href{http://ascl.net/#1}{\nolinkurl{http://ascl.net/#1}}}
\providecommand{\doarXiv}[1]{\href{https://arxiv.org/abs/#1}{\nolinkurl{https://arxiv.org/abs/#1}}}

\bibitem[{{Abdo} {et~al.}(2010){Abdo}, {Ackermann}, {Ajello}, {Allafort},
  {Antolini}, {Atwood}, {Axelsson}, {Baldini}, {Ballet}, {Barbiellini},
  {Bastieri}, {Baughman}, {Bechtol}, {Bellazzini}, {Belli}, {Berenji},
  {Bisello}, {Blandford}, {Bloom}, {Bonamente}, {Bonnell}, {Borgland},
  {Bouvier}, {Bregeon}, {Brez}, {Brigida}, {Bruel}, {Burnett}, {Busetto},
  {Buson}, {Caliandro}, {Cameron}, {Campana}, {Canadas}, {Caraveo}, {Carrigan},
  {Casandjian}, {Cavazzuti}, {Ceccanti}, {Cecchi}, {{\c{C}}elik}, {Charles},
  {Chekhtman}, {Cheung}, {Chiang}, {Cillis}, {Ciprini}, {Claus},
  {Cohen-Tanugi}, {Conrad}, {Corbet}, {Davis}, {DeKlotz}, {den Hartog},
  {Dermer}, {de Angelis}, {de Luca}, {de Palma}, {Digel}, {Dormody}, {Silva},
  {Drell}, {Dubois}, {Dumora}, {Fabiani}, {Farnier}, {Favuzzi}, {Fegan},
  {Ferrara}, {Focke}, {Fortin}, {Frailis}, {Fukazawa}, {Funk}, {Fusco},
  {Gargano}, {Gasparrini}, {Gehrels}, {Germani}, {Giavitto}, {Giebels},
  {Giglietto}, {Giommi}, {Giordano}, {Giroletti}, {Glanzman}, {Godfrey},
  {Grenier}, {Grondin}, {Grove}, {Guillemot}, {Guiriec}, {Gustafsson},
  {Hadasch}, {Hanabata}, {Harding}, {Hayashida}, {Hays}, {Healey}, {Hill},
  {Horan}, {Hughes}, {Iafrate}, {J{\'o}hannesson}, {Johnson}, {Johnson},
  {Johnson}, {Johnson}, {Kamae}, {Katagiri}, {Kataoka}, {Kawai}, {Kerr},
  {Kn{\"o}dlseder}, {Kocevski}, {Kuss}, {Lande}, {Landriu}, {Latronico}, {Lee},
  {Lemoine-Goumard}, {Lionetto}, {Llena Garde}, {Longo}, {Loparco}, {Lott},
  {Lovellette}, {Lubrano}, {Madejski}, {Makeev}, {Marangelli}, {Marelli},
  {Massaro}, {Mazziotta}, {McConville}, {McEnery}, {Michelson}, {Minuti},
  {Mitthumsiri}, {Mizuno}, {Moiseev}, {Mongelli}, {Monte}, {Monzani},
  {Moretti}, {Morselli}, {Moskalenko}, {Murgia}, {Nakajima}, {Nakamori},
  {Naumann-Godo}, {Nolan}, {Norris}, {Nuss}, {Ohno}, {Ohsugi}, {Omodei},
  {Orlando}, {Ormes}, {Ozaki}, {Paccagnella}, {Paneque}, {Panetta}, {Parent},
  {Pelassa}, {Pepe}, {Pesce-Rollins}, {Pinchera}, {Piron}, {Porter}, {Poupard},
  {Rain{\`o}}, {Rando}, {Ray}, {Razzano}, {Razzaque}, {Rea}, {Reimer},
  {Reimer}, {Reposeur}, {Ripken}, {Ritz}, {Rochester}, {Rodriguez}, {Romani},
  {Roth}, {Sadrozinski}, {Salvetti}, {Sanchez}, {Sander}, {Saz Parkinson},
  {Scargle}, {Schalk}, {Scolieri}, {Sgr{\`o}}, {Shaw}, {Siskind}, {Smith},
  {Smith}, {Spandre}, {Spinelli}, {Starck}, {Stephens}, {Striani}, {Strickman},
  {Strong}, {Suson}, {Tajima}, {Takahashi}, {Takahashi}, {Tanaka}, {Thayer},
  {Thayer}, {Thompson}, {Tibaldo}, {Tibolla}, {Tinebra}, {Torres}, {Tosti},
  {Tramacere}, {Uchiyama}, {Usher}, {Van Etten}, {Vasileiou}, {Vilchez},
  {Vitale}, {Waite}, {Wallace}, {Wang}, {Watters}, {Winer}, {Wood}, {Yang},
  {Ylinen}, {Ziegler}, \& {Fermi LAT Collaboration}}]{1FGL_2010}
{Abdo}, A.~A., {Ackermann}, M., {Ajello}, M., {et~al.} 2010, \apjs, 188, 405,
  \dodoi{10.1088/0067-0049/188/2/405}

\bibitem[{{Abdollahi} {et~al.}(2022){Abdollahi}, {Acero}, {Baldini}, {Ballet},
  {Bastieri}, {Bellazzini}, {Berenji}, {Berretta}, {Bissaldi}, {Blandford},
  {Bloom}, {Bonino}, {Brill}, {Britto}, {Bruel}, {Burnett}, {Buson}, {Cameron},
  {Caputo}, {Caraveo}, {Castro}, {Chaty}, {Cheung}, {Chiaro}, {Cibrario},
  {Ciprini}, {Coronado-Bl{\'a}zquez}, {Crnogorcevic}, {Cutini}, {D'Ammando},
  {De Gaetano}, {Digel}, {Di Lalla}, {Dirirsa}, {Di Venere}, {Dom{\'\i}nguez},
  {Fallah Ramazani}, {Fegan}, {Ferrara}, {Fiori}, {Fleischhack}, {Franckowiak},
  {Fukazawa}, {Funk}, {Fusco}, {Galanti}, {Gammaldi}, {Gargano}, {Garrappa},
  {Gasparrini}, {Giacchino}, {Giglietto}, {Giordano}, {Giroletti}, {Glanzman},
  {Green}, {Grenier}, {Grondin}, {Guillemot}, {Guiriec}, {Gustafsson},
  {Harding}, {Hays}, {Hewitt}, {Horan}, {Hou}, {J{\'o}hannesson}, {Karwin},
  {Kayanoki}, {Kerr}, {Kuss}, {Landriu}, {Larsson}, {Latronico},
  {Lemoine-Goumard}, {Li}, {Liodakis}, {Longo}, {Loparco}, {Lott}, {Lubrano},
  {Maldera}, {Malyshev}, {Manfreda}, {Mart{\'\i}-Devesa}, {Mazziotta}, {Mereu},
  {Meyer}, {Michelson}, {Mirabal}, {Mitthumsiri}, {Mizuno}, {Moiseev},
  {Monzani}, {Morselli}, {Moskalenko}, {Negro}, {Nuss}, {Omodei}, {Orienti},
  {Orlando}, {Paneque}, {Pei}, {Perkins}, {Persic}, {Pesce-Rollins},
  {Petrosian}, {Pillera}, {Poon}, {Porter}, {Principe}, {Rain{\`o}}, {Rando},
  {Rani}, {Razzano}, {Razzaque}, {Reimer}, {Reimer}, {Reposeur},
  {S{\'a}nchez-Conde}, {Saz Parkinson}, {Scotton}, {Serini}, {Sgr{\`o}},
  {Siskind}, {Smith}, {Spandre}, {Spinelli}, {Sueoka}, {Suson}, {Tajima},
  {Tak}, {Thayer}, {Thompson}, {Torres}, {Troja}, {Valverde}, {Wood}, \&
  {Zaharijas}}]{4FGL_2022}
{Abdollahi}, S., {Acero}, F., {Baldini}, L., {et~al.} 2022, \apjs, 260, 53,
  \dodoi{10.3847/1538-4365/ac6751}

\bibitem[{{Acero} {et~al.}(2015){Acero}, {Ackermann}, {Ajello}, {Albert},
  {Atwood}, {Axelsson}, {Baldini}, {Ballet}, {Barbiellini}, {Bastieri},
  {Belfiore}, {Bellazzini}, {Bissaldi}, {Blandford}, {Bloom}, {Bogart},
  {Bonino}, {Bottacini}, {Bregeon}, {Britto}, {Bruel}, {Buehler}, {Burnett},
  {Buson}, {Caliandro}, {Cameron}, {Caputo}, {Caragiulo}, {Caraveo},
  {Casandjian}, {Cavazzuti}, {Charles}, {Chaves}, {Chekhtman}, {Cheung},
  {Chiang}, {Chiaro}, {Ciprini}, {Claus}, {Cohen-Tanugi}, {Cominsky}, {Conrad},
  {Cutini}, {D'Ammando}, {de Angelis}, {DeKlotz}, {de Palma}, {Desiante},
  {Digel}, {Di Venere}, {Drell}, {Dubois}, {Dumora}, {Favuzzi}, {Fegan},
  {Ferrara}, {Finke}, {Franckowiak}, {Fukazawa}, {Funk}, {Fusco}, {Gargano},
  {Gasparrini}, {Giebels}, {Giglietto}, {Giommi}, {Giordano}, {Giroletti},
  {Glanzman}, {Godfrey}, {Grenier}, {Grondin}, {Grove}, {Guillemot}, {Guiriec},
  {Hadasch}, {Harding}, {Hays}, {Hewitt}, {Hill}, {Horan}, {Iafrate}, {Jogler},
  {J{\'o}hannesson}, {Johnson}, {Johnson}, {Johnson}, {Johnson}, {Kamae},
  {Kataoka}, {Katsuta}, {Kuss}, {La Mura}, {Landriu}, {Larsson}, {Latronico},
  {Lemoine-Goumard}, {Li}, {Li}, {Longo}, {Loparco}, {Lott}, {Lovellette},
  {Lubrano}, {Madejski}, {Massaro}, {Mayer}, {Mazziotta}, {McEnery},
  {Michelson}, {Mirabal}, {Mizuno}, {Moiseev}, {Mongelli}, {Monzani},
  {Morselli}, {Moskalenko}, {Murgia}, {Nuss}, {Ohno}, {Ohsugi}, {Omodei},
  {Orienti}, {Orlando}, {Ormes}, {Paneque}, {Panetta}, {Perkins},
  {Pesce-Rollins}, {Piron}, {Pivato}, {Porter}, {Racusin}, {Rando}, {Razzano},
  {Razzaque}, {Reimer}, {Reimer}, {Reposeur}, {Rochester}, {Romani},
  {Salvetti}, {S{\'a}nchez-Conde}, {Saz Parkinson}, {Schulz}, {Siskind},
  {Smith}, {Spada}, {Spandre}, {Spinelli}, {Stephens}, {Strong}, {Suson},
  {Takahashi}, {Takahashi}, {Tanaka}, {Thayer}, {Thayer}, {Thompson},
  {Tibaldo}, {Tibolla}, {Torres}, {Torresi}, {Tosti}, {Troja}, {Van Klaveren},
  {Vianello}, {Winer}, {Wood}, {Wood}, {Zimmer}, \& {Fermi-LAT
  Collaboration}}]{3FGL_2015}
{Acero}, F., {Ackermann}, M., {Ajello}, M., {et~al.} 2015, \apjs, 218, 23,
  \dodoi{10.1088/0067-0049/218/2/23}

\bibitem[{{Ackermann} {et~al.}(2012){Ackermann}, {Ajello}, {Allafort},
  {Schady}, {Baldini}, {Ballet}, {Barbiellini}, {Bastieri}, {Bellazzini},
  {Blandford}, {Bloom}, {Borgland}, {Bottacini}, {Bouvier}, {Bregeon},
  {Brigida}, {Bruel}, {Buehler}, {Buson}, {Caliandro}, {Cameron}, {Caraveo},
  {Cavazzuti}, {Cecchi}, {Charles}, {Chaves}, {Chekhtman}, {Cheung}, {Chiang},
  {Chiaro}, {Ciprini}, {Claus}, {Cohen-Tanugi}, {Conrad}, {Cutini},
  {D'Ammando}, {de Palma}, {Dermer}, {Digel}, {do Couto e Silva},
  {Dom{\'{\i}}nguez}, {Drell}, {Drlica-Wagner}, {Favuzzi}, {Fegan}, {Focke},
  {Franckowiak}, {Fukazawa}, {Funk}, {Fusco}, {Gargano}, {Gasparrini},
  {Gehrels}, {Germani}, {Giglietto}, {Giordano}, {Giroletti}, {Glanzman},
  {Godfrey}, {Grenier}, {Grove}, {Guiriec}, {Gustafsson}, {Hadasch},
  {Hayashida}, {Hays}, {Jackson}, {Jogler}, {Kataoka}, {Kn{\"o}dlseder},
  {Kuss}, {Lande}, {Larsson}, {Latronico}, {Longo}, {Loparco}, {Lovellette},
  {Lubrano}, {Mazziotta}, {McEnery}, {Mehault}, {Michelson}, {Mizuno}, {Monte},
  {Monzani}, {Morselli}, {Moskalenko}, {Murgia}, {Tramacere}, {Nuss},
  {Greiner}, {Ohno}, {Ohsugi}, {Omodei}, {Orienti}, {Orlando}, {Ormes},
  {Paneque}, {Perkins}, {Pesce-Rollins}, {Piron}, {Pivato}, {Porter},
  {Rain{\`o}}, {Rando}, {Razzano}, {Razzaque}, {Reimer}, {Reimer}, {Reyes},
  {Ritz}, {Rau}, {Romoli}, {Roth}, {S{\'a}nchez-Conde}, {Sanchez}, {Scargle},
  {Sgr{\`o}}, {Siskind}, {Spandre}, {Spinelli}, {Stawarz}, {Suson},
  {Takahashi}, {Tanaka}, {Thayer}, {Thompson}, {Tibaldo}, {Tinivella},
  {Torres}, {Tosti}, {Troja}, {Usher}, {Vandenbroucke}, {Vasileiou},
  {Vianello}, {Vitale}, {Waite}, {Winer}, {Wood}, \& {Wood}}]{ackermann12}
{Ackermann}, M., {Ajello}, M., {Allafort}, A., {et~al.} 2012, Science, 338,
  1190, \dodoi{10.1126/science.1227160}

\bibitem[{{Ajello} {et~al.}(2012){Ajello}, {Shaw}, {Romani}, {Dermer},
  {Costamante}, {King}, {Max-Moerbeck}, {Readhead}, {Reimer}, {Richards}, \&
  {Stevenson}}]{ajello12}
{Ajello}, M., {Shaw}, M.~S., {Romani}, R.~W., {et~al.} 2012, \apj, 751, 108,
  \dodoi{10.1088/0004-637X/751/2/108}

\bibitem[{{Ajello} {et~al.}(2014){Ajello}, {Romani}, {Gasparrini}, {Shaw},
  {Bolmer}, {Cotter}, {Finke}, {Greiner}, {Healey}, {King}, {Max-Moerbeck},
  {Michelson}, {Potter}, {Rau}, {Readhead}, {Richards}, \& {Schady}}]{ajello14}
{Ajello}, M., {Romani}, R.~W., {Gasparrini}, D., {et~al.} 2014, \apj, 780, 73,
  \dodoi{10.1088/0004-637X/780/1/73}

\bibitem[{{Ajello} {et~al.}(2017){Ajello}, {Atwood}, {Baldini}, {Ballet},
  {Barbiellini}, {Bastieri}, {Bellazzini}, {Bissaldi}, {Blandford}, {Bloom},
  {Bonino}, {Bregeon}, {Britto}, {Bruel}, {Buehler}, {Buson}, {Cameron},
  {Caputo}, {Caragiulo}, {Caraveo}, {Cavazzuti}, {Cecchi}, {Charles},
  {Chekhtman}, {Cheung}, {Chiaro}, {Ciprini}, {Cohen}, {Costantin}, {Costanza},
  {Cuoco}, {Cutini}, {D'Ammando}, {de Palma}, {Desiante}, {Digel}, {Di Lalla},
  {Di Mauro}, {Di Venere}, {Dom{\'{\i}}nguez}, {Drell}, {Dumora}, {Favuzzi},
  {Fegan}, {Ferrara}, {Fortin}, {Franckowiak}, {Fukazawa}, {Funk}, {Fusco},
  {Gargano}, {Gasparrini}, {Giglietto}, {Giommi}, {Giordano}, {Giroletti},
  {Glanzman}, {Green}, {Grenier}, {Grondin}, {Grove}, {Guillemot}, {Guiriec},
  {Harding}, {Hays}, {Hewitt}, {Horan}, {J{\'o}hannesson}, {Kensei}, {Kuss},
  {La Mura}, {Larsson}, {Latronico}, {Lemoine-Goumard}, {Li}, {Longo},
  {Loparco}, {Lott}, {Lubrano}, {Magill}, {Maldera}, {Manfreda}, {Mazziotta},
  {McEnery}, {Meyer}, {Michelson}, {Mirabal}, {Mitthumsiri}, {Mizuno},
  {Moiseev}, {Monzani}, {Morselli}, {Moskalenko}, {Negro}, {Nuss}, {Ohsugi},
  {Omodei}, {Orienti}, {Orlando}, {Palatiello}, {Paliya}, {Paneque}, {Perkins},
  {Persic}, {Pesce-Rollins}, {Piron}, {Porter}, {Principe}, {Rain{\`o}},
  {Rando}, {Razzano}, {Razzaque}, {Reimer}, {Reimer}, {Reposeur}, {Saz
  Parkinson}, {Sgr{\`o}}, {Simone}, {Siskind}, {Spada}, {Spandre}, {Spinelli},
  {Stawarz}, {Suson}, {Takahashi}, {Tak}, {Thayer}, {Thayer}, {Thompson},
  {Torres}, {Torresi}, {Troja}, {Vianello}, {Wood}, \& {Wood}}]{ajello17}
{Ajello}, M., {Atwood}, W.~B., {Baldini}, L., {et~al.} 2017, \apjs, 232, 18,
  \dodoi{10.3847/1538-4365/aa8221}

\bibitem[{{{\'A}lvarez Crespo} {et~al.}(2016{\natexlab{a}}){{\'A}lvarez
  Crespo}, {Masetti}, {Ricci}, {Landoni}, {Pati{\~n}o-{\'A}lvarez}, {Massaro},
  {D'Abrusco}, {Paggi}, {Chavushyan}, {Jim{\'e}nez-Bail{\'o}n}, {Torrealba},
  {Latronico}, {La Franca}, {Smith}, \& {Tosti}}]{alvarez16a}
{{\'A}lvarez Crespo}, N., {Masetti}, N., {Ricci}, F., {et~al.}
  2016{\natexlab{a}}, \aj, 151, 32, \dodoi{10.3847/0004-6256/151/2/32}

\bibitem[{{{\'A}lvarez Crespo} {et~al.}(2016{\natexlab{b}}){{\'A}lvarez
  Crespo}, {Massaro}, {Milisavljevic}, {Landoni}, {Chavushyan},
  {Pati{\~n}o-{\'A}lvarez}, {Masetti}, {Jim{\'e}nez-Bail{\'o}n}, {Strader},
  {Chomiuk}, {Katagiri}, {Kagaya}, {Cheung}, {Paggi}, {D'Abrusco}, {Ricci}, {La
  Franca}, {Smith}, \& {Tosti}}]{alvarez16b}
{{\'A}lvarez Crespo}, N., {Massaro}, F., {Milisavljevic}, D., {et~al.}
  2016{\natexlab{b}}, \aj, 151, 95, \dodoi{10.3847/0004-6256/151/4/95}

\bibitem[{{Archer} {et~al.}(2018){Archer}, {Benbow}, {Bird}, {Brose},
  {Buchovecky}, {Bugaev}, {Cui}, {Daniel}, {Falcone}, {Feng}, {Finley},
  {Flinders}, {Fortson}, {Furniss}, {Gillanders}, {H{\"u}tten}, {Hanna},
  {Hervet}, {Holder}, {Hughes}, {Humensky}, {Johnson}, {Kaaret}, {Kar},
  {Kelley-Hoskins}, {Kieda}, {Krause}, {Krennrich}, {Kumar}, {Lang}, {Lin},
  {McArthur}, {Moriarty}, {Mukherjee}, {Nieto}, {O'Brien}, {Ong}, {Otte},
  {Park}, {Petrashyk}, {Pohl}, {Popkow}, {Pueschel}, {Quinn}, {Ragan},
  {Reynolds}, {Richards}, {Roache}, {Rulten}, {Sadeh}, {Sembroski},
  {Shahinyan}, {Tyler}, {Wakely}, {Weiner}, {Weinstein}, {Wells}, {Wilcox},
  {Wilhelm}, {Williams}, {VERITAS Collaboration}, {Brisken}, \&
  {Pontrelli}}]{Archer_2018}
{Archer}, A., {Benbow}, W., {Bird}, R., {et~al.} 2018, \apj, 862, 41,
  \dodoi{10.3847/1538-4357/aacbd0}

\bibitem[{{Atwood} {et~al.}(2009){Atwood}, {Abdo}, {Ackermann}, {Althouse},
  {Anderson}, {Axelsson}, {Baldini}, {Ballet}, {Band}, {Barbiellini}, \&
  et~al.}]{atwood09}
{Atwood}, W.~B., {Abdo}, A.~A., {Ackermann}, M., {et~al.} 2009, \apj, 697,
  1071, \dodoi{10.1088/0004-637X/697/2/1071}

\bibitem[{{Celotti} \& {Ghisellini}(2008)}]{Celotti_2008}
{Celotti}, A., \& {Ghisellini}, G. 2008, \mnras, 385, 283,
  \dodoi{10.1111/j.1365-2966.2007.12758.x}

\bibitem[{{Costamante} {et~al.}(2001){Costamante}, {Ghisellini}, {Giommi},
  {Tagliaferri}, {Celotti}, {Chiaberge}, {Fossati}, {Maraschi}, {Tavecchio},
  {Treves}, \& {Wolter}}]{Costamante_2001}
{Costamante}, L., {Ghisellini}, G., {Giommi}, P., {et~al.} 2001, \aap, 371,
  512, \dodoi{10.1051/0004-6361:20010412}

\bibitem[{{de Menezes} {et~al.}(2020){de Menezes}, {Amaya-Almaz{\'a}n},
  {Marchesini}, {Pe{\~n}a-Herazo}, {Massaro}, {Chavushyan}, {Paggi}, {Landoni},
  {Masetti}, {Ricci}, {D'Abrusco}, {La Franca}, {Smith}, {Milisavljevic},
  {Tosti}, {Jim{\'e}nez-Bail{\'o}n}, \& {Cheung}}]{deMenezes_2020}
{de Menezes}, R., {Amaya-Almaz{\'a}n}, R.~A., {Marchesini}, E.~J., {et~al.}
  2020, \apss, 365, 12, \dodoi{10.1007/s10509-020-3727-5}

\bibitem[{{Desai} {et~al.}(2019){Desai}, {Helgason}, {Ajello}, {Paliya},
  {Dom{\'\i}nguez}, {Finke}, \& {Hartmann}}]{Deasi_EBL_2019}
{Desai}, A., {Helgason}, K., {Ajello}, M., {et~al.} 2019, \apjl, 874, L7,
  \dodoi{10.3847/2041-8213/ab0c10}

\bibitem[{Desai {et~al.}(2019)Desai, Marchesi, Rajagopal, \&
  Ajello}]{desai2019}
Desai, A., Marchesi, S., Rajagopal, M., \& Ajello, M. 2019, \apss, 241, 5,
  \dodoi{10.3847/1538-4365/ab01fc}

\bibitem[{{Dom{\'{\i}}nguez} {et~al.}(2013){Dom{\'{\i}}nguez}, {Finke},
  {Prada}, {Primack}, {Kitaura}, {Siana}, \& {Paneque}}]{dominguez13}
{Dom{\'{\i}}nguez}, A., {Finke}, J.~D., {Prada}, F., {et~al.} 2013, \apj, 770,
  77, \dodoi{10.1088/0004-637X/770/1/77}

\bibitem[{{Fermi-LAT Collaboration} {et~al.}(2018){Fermi-LAT Collaboration},
  {Abdollahi}, {Ackermann}, {Ajello}, {Atwood}, {Baldini}, {Ballet},
  {Barbiellini}, {Bastieri}, {Becerra Gonzalez}, {Bellazzini}, {Bissaldi},
  {Blandford}, {Bloom}, {Bonino}, {Bottacini}, {Buson}, {Bregeon}, {Bruel},
  {Buehler}, {Cameron}, {Caputo}, {Caraveo}, {Cavazzuti}, {Charles}, {Chen},
  {Cheung}, {Chiaro}, {Ciprini}, {Cohen-Tanugi}, {Cominsky}, {Conrad},
  {Costantin}, {Cutini}, {D'Ammando}, {de Palma}, {Desai}, {Digel}, {Di Lalla},
  {Di Mauro}, {Di Venere}, {Dom{\'\i}nguez}, {Favuzzi}, {Fegan}, {Finke},
  {Franckowiak}, {Fukazawa}, {Funk}, {Fusco}, {Gallardo Romero}, {Gargano},
  {Gasparrini}, {Giglietto}, {Giordano}, {Giroletti}, {Green}, {Grenier},
  {Guillemot}, {Guiriec}, {Hartmann}, {Hays}, {Helgason}, {Horan},
  {J{\'o}hannesson}, {Kocevski}, {Kuss}, {Larsson}, {Latronico}, {Li}, {Longo},
  {Loparco}, {Lott}, {Lovellette}, {Lubrano}, {Madejski}, {Magill}, {Maldera},
  {Manfreda}, {Marcotulli}, {Mazziotta}, {McEnery}, {Meyer}, {Michelson},
  {Mizuno}, {Monzani}, {Morselli}, {Moskalenko}, {Negro}, {Nuss}, {Ojha},
  {Omodei}, {Orienti}, {Orlando}, {Ormes}, {Palatiello}, {Paliya}, {Paneque},
  {Perkins}, {Persic}, {Pesce-Rollins}, {Petrosian}, {Piron}, {Porter},
  {Primack}, {Principe}, {Rain{\`o}}, {Rando}, {Razzano}, {Razzaque}, {Reimer},
  {Reimer}, {Saz Parkinson}, {Sgr{\`o}}, {Siskind}, {Spandre}, {Spinelli},
  {Suson}, {Tajima}, {Takahashi}, {Thayer}, {Tibaldo}, {Torres}, {Torresi},
  {Tosti}, {Tramacere}, {Troja}, {Valverde}, {Vianello}, {Vogel}, {Wood}, \&
  {Zaharijas}}]{EBL_2018}
{Fermi-LAT Collaboration}, {Abdollahi}, S., {Ackermann}, M., {et~al.} 2018,
  Science, 362, 1031, \dodoi{10.1126/science.aat8123}

\bibitem[{{Frank} {et~al.}(2002){Frank}, {King}, \& {Raine}}]{Frank_2002}
{Frank}, J., {King}, A., \& {Raine}, D.~J. 2002, {Accretion Power in
  Astrophysics, by Juhan Frank and Andrew King and Derek Raine, pp.~398.~ISBN
  0521620538.~Cambridge, UK: Cambridge University Press, February 2002}

\bibitem[{{Ghisellini} {et~al.}(2017){Ghisellini}, {Righi}, {Costamante}, \&
  {Tavecchio}}]{ghisellini17}
{Ghisellini}, G., {Righi}, C., {Costamante}, L., \& {Tavecchio}, F. 2017,
  \mnras, 469, 255, \dodoi{10.1093/mnras/stx806}

\bibitem[{{Ghisellini} \& {Tavecchio}(2009)}]{Ghisellini_2009}
{Ghisellini}, G., \& {Tavecchio}, F. 2009, \mnras, 397, 985,
  \dodoi{10.1111/j.1365-2966.2009.15007.x}

\bibitem[{{Ghisellini} {et~al.}(2012){Ghisellini}, {Tavecchio}, {Foschini},
  {Sbarrato}, {Ghirlanda}, \& {Maraschi}}]{ghisellini2012blue}
{Ghisellini}, G., {Tavecchio}, F., {Foschini}, L., {et~al.} 2012, \mnras, 425,
  1371, \dodoi{10.1111/j.1365-2966.2012.21554.x}

\bibitem[{Giommi {et~al.}(2002)Giommi, Padovani, Perri, Landt, \&
  Perlman}]{Giommi2002}
Giommi, P., Padovani, P., Perri, M., Landt, H., \& Perlman, E. 2002.
\newblock \doarXiv{astro-ph/0209621}

\bibitem[{{Goldoni} {et~al.}(2021){Goldoni}, {Pita}, {Boisson}, {Max-Moerbeck},
  {Kasai}, {Williams}, {D'Ammando}, {Navarro-Aranguiz}, {Backes}, {Barres de
  Almeida}, {Becerra-Gonzalez}, {Cotter}, {Hervet}, {Lenain}, {Lindfors},
  {Sol}, \& {Wagner}}]{Goldoni_2021}
{Goldoni}, P., {Pita}, S., {Boisson}, C., {et~al.} 2021, \aap, 650, A106,
  \dodoi{10.1051/0004-6361/202040090}

\bibitem[{{IceCube Collaboration} {et~al.}(2018){IceCube Collaboration},
  {Aartsen}, {Ackermann}, {Adams}, {Aguilar}, {Ahlers}, {Ahrens}, {Samarai},
  {Altmann}, {Andeen}, {Anderson}, {Ansseau}, {Anton}, {Arg{\"u}elles},
  {Arsioli}, {Auffenberg}, {Axani}, {Bagherpour}, {Bai}, {Barron}, {Barwick},
  {Baum}, {Bay}, {Beatty}, {Becker Tjus}, {Becker}, {BenZvi}, {Berley},
  {Bernardini}, {Besson}, {Binder}, {Bindig}, {Blaufuss}, {Blot}, {Bohm},
  {B{\"o}rner}, {Bos}, {B{\"o}ser}, {Botner}, {Bourbeau}, {Bourbeau},
  {Bradascio}, {Braun}, {Brenzke}, {Bretz}, {Bron}, {Brostean-Kaiser},
  {Burgman}, {Busse}, {Carver}, {Cheung}, {Chirkin}, {Christov}, {Clark},
  {Classen}, {Coenders}, {Collin}, {Conrad}, {Coppin}, {Correa}, {Cowen},
  {Cross}, {Dave}, {Day}, {de Andr{\'e}}, {De Clercq}, {DeLaunay}, {Dembinski},
  {DeRidder}, {Desiati}, {de Vries}, {de Wasseige}, {de With}, {DeYoung},
  {D{\'\i}az-V{\'e}lez}, {di Lorenzo}, {Dujmovic}, {Dumm}, {Dunkman}, {Dvorak},
  {Eberhardt}, {Ehrhardt}, {Eichmann}, {Eller}, {Evenson}, {Fahey}, {Fazely},
  {Felde}, {Filimonov}, {Finley}, {Flis}, {Franckowiak}, {Friedman}, {Fritz},
  {Gaisser}, {Gallagher}, {Gerhardt}, {Ghorbani}, {Giommi}, {Glauch},
  {Gl{\"u}senkamp}, {Goldschmidt}, {Gonzalez}, {Grant}, {Griffith}, {Haack},
  {Hallgren}, {Halzen}, {Hanson}, {Hebecker}, {Heereman}, {Helbing},
  {Hellauer}, {Hickford}, {Hignight}, {Hill}, {Hoffman}, {Hoffmann}, {Hoinka},
  {Hokanson-Fasig}, {Hoshina}, {Huang}, {Huber}, {Hultqvist}, {H{\"u}nnefeld},
  {Hussain}, {In}, {Iovine}, {Ishihara}, {Jacobi}, {Japaridze}, {Jeong},
  {Jero}, {Jones}, {Kalaczynski}, {Kang}, {Kappes}, {Kappesser}, {Karg},
  {Karle}, {Katz}, {Kauer}, {Keivani}, {Kelley}, {Kheirandish}, {Kim}, {Kim},
  {Kintscher}, {Kiryluk}, {Kittler}, {Klein}, {Koirala}, {Kolanoski},
  {K{\"o}pke}, {Kopper}, {Kopper}, {Koschinsky}, {Koskinen}, {Kowalski},
  {Krammer}, {Krings}, {Kroll}, {Kr{\"u}ckl}, {Kunwar}, {Kurahashi},
  {Kuwabara}, {Kyriacou}, {Labare}, {Lanfranchi}, {Larson}, {Lauber},
  {Leonard}, {Lesiak-Bzdak}, {Leuermann}, {Liu}, {Lozano Mariscal}, {Lu},
  {L{\"u}nemann}, {Luszczak}, {Madsen}, {Maggi}, {Mahn}, {Mancina}, {Maruyama},
  {Mase}, {Maunu}, {Meagher}, {Medici}, {Meier}, {Menne}, {Merino}, {Meures},
  {Miarecki}, {Micallef}, {Moment{\'e}}, {Montaruli}, {Moore}, {Morse},
  {Moulai}, {Nahnhauer}, {Nakarmi}, {Naumann}, {Neer}, {Niederhausen},
  {Nowicki}, {Nygren}, {Obertacke Pollmann}, {Olivas}, {O'Murchadha},
  {O'Sullivan}, {Padovani}, {Palczewski}, {Pandya}, {Pankova}, {Peiffer},
  {Pepper}, {P{\'e}rez de los Heros}, {Pieloth}, {Pinat}, {Plum}, {Price},
  {Przybylski}, {Raab}, {R{\"a}del}, {Rameez}, {Rawlins}, {Rea}, {Reimann},
  {Relethford}, {Relich}, {Resconi}, {Rhode}, {Richman}, {Robertson}, {Rongen},
  {Rott}, {Ruhe}, {Ryckbosch}, {Rysewyk}, {Safa}, {Sahakyan}, {S{\"a}lzer},
  {Sanchez Herrera}, {Sandrock}, {Sandroos}, {Santander}, {Sarkar}, {Sarkar},
  {Satalecka}, {Schlunder}, {Schmidt}, {Schneider}, {Schoenen},
  {Sch{\"o}neberg}, {Schumacher}, {Sclafani}, {Seckel}, {Seunarine},
  {Soedingrekso}, {Soldin}, {Song}, {Spiczak}, {Spiering}, {Stachurska},
  {Stamatikos}, {Stanev}, {Stasik}, {Stettner}, {Steuer}, {Stezelberger},
  {Stokstad}, {St{\"o}{\ss}l}, {Strotjohann}, {Stuttard}, {Sullivan},
  {Sutherland}, {Taboada}, {Tatar}, {Tenholt}, {Ter-Antonyan}, {Terliuk},
  {Tilav}, {Toale}, {Tobin}, {Toennis}, {Toscano}, {Tosi}, {Tselengidou},
  {Tung}, {Turcati}, {Turley}, {Ty}, {Unger}, {Usner}, {Vandenbroucke}, {Van
  Driessche}, {van Eijk}, {van Eijndhoven}, {Vanheule}, {van Santen}, {Vogel},
  {Vraeghe}, {Walck}, {Wallace}, {Wallraff}, {Wandler}, {Wandkowsky}, {Waza},
  {Weaver}, {Weiss}, {Wendt}, {Werthebach}, {Westerhoff}, {Whelan},
  {Whitehorn}, {Wiebe}, {Wiebusch}, {Wille}, {Williams}, {Wills}, {Wolf},
  {Wood}, {Wood}, {Woschnagg}, {Xu}, {Xu}, {Xu}, {Yanez}, {Yodh}, {Yoshida}, \&
  {Yuan}}]{IceCube_2018}
{IceCube Collaboration}, {Aartsen}, M.~G., {Ackermann}, M., {et~al.} 2018,
  Science, 361, 147, \dodoi{10.1126/science.aat2890}

\bibitem[{{Joffre} {et~al.}(2022){Joffre}, {Silver}, {Rajagopal}, {Ajello},
  {Torres-Alb{\`a}}, {Pizzetti}, {Marchesi}, \& {Kaur}}]{Joffre_2022}
{Joffre}, S., {Silver}, R., {Rajagopal}, M., {et~al.} 2022, \apj, 940, 139,
  \dodoi{10.3847/1538-4357/ac9797}

\bibitem[{{Kalberla} {et~al.}(2005){Kalberla}, {Burton}, {Hartmann}, {Arnal},
  {Bajaja}, {Morras}, \& {P{\"o}ppel}}]{Kaberla_2005}
{Kalberla}, P.~M.~W., {Burton}, W.~B., {Hartmann}, D., {et~al.} 2005, \aap,
  440, 775, \dodoi{10.1051/0004-6361:20041864}

\bibitem[{{Kaur} {et~al.}(2019){Kaur}, {Ajello}, {Marchesi}, \&
  {Omodei}}]{Kaur_2019}
{Kaur}, A., {Ajello}, M., {Marchesi}, S., \& {Omodei}, N. 2019, \apj, 871, 94,
  \dodoi{10.3847/1538-4357/aaf649}

\bibitem[{{King} {et~al.}(2016){King}, {Miller}, {Bietenholz}, {G{\"u}ltekin},
  {Reynolds}, {Mioduszewski}, {Rupen}, \& {Bartel}}]{King_2016}
{King}, A.~L., {Miller}, J.~M., {Bietenholz}, M., {et~al.} 2016, Nature
  Physics, 12, 772, \dodoi{10.1038/nphys3724}

\bibitem[{{Landoni} {et~al.}(2015){Landoni}, {Massaro}, {Paggi}, {D'Abrusco},
  {Milisavljevic}, {Masetti}, {Smith}, {Tosti}, {Chomiuk}, {Strader}, \&
  {Cheung}}]{landoni15}
{Landoni}, M., {Massaro}, F., {Paggi}, A., {et~al.} 2015, \aj, 149, 163,
  \dodoi{10.1088/0004-6256/149/5/163}

\bibitem[{{Marchesi} {et~al.}(2018){Marchesi}, {Kaur}, \&
  {Ajello}}]{marchesi18}
{Marchesi}, S., {Kaur}, A., \& {Ajello}, M. 2018, \aj, 156, 212,
  \dodoi{10.3847/1538-3881/aae201}

\bibitem[{{Marchesini} {et~al.}(2016){Marchesini}, {Masetti}, {Chavushyan},
  {Cellone}, {Andruchow}, {Bassani}, {Bazzano}, {Jim{\'e}nez-Bail{\'o}n},
  {Landi}, {Malizia}, {Palazzi}, {Pati{\~n}o-{\'A}lvarez},
  {Rodr{\'{\i}}guez-Castillo}, {Stephen}, \& {Ubertini}}]{marchesini16}
{Marchesini}, E.~J., {Masetti}, N., {Chavushyan}, V., {et~al.} 2016, \aap, 596,
  A10, \dodoi{10.1051/0004-6361/201629028}

\bibitem[{{Massaro} {et~al.}(2014){Massaro}, {Masetti}, {D'Abrusco}, {Paggi},
  \& {Funk}}]{massaro14}
{Massaro}, F., {Masetti}, N., {D'Abrusco}, R., {Paggi}, A., \& {Funk}, S. 2014,
  \aj, 148, 66, \dodoi{10.1088/0004-6256/148/4/66}

\bibitem[{{Massaro} {et~al.}(2016){Massaro}, {{\'A}lvarez Crespo}, {D'Abrusco},
  {Landoni}, {Masetti}, {Ricci}, {Milisavljevic}, {Paggi}, {Chavushyan},
  {Jim{\'e}nez-Bail{\'o}n}, {Pati{\~n}o-{\'A}lvarez}, {Strader}, {Chomiuk}, {La
  Franca}, {Smith}, \& {Tosti}}]{Massaro_2016}
{Massaro}, F., {{\'A}lvarez Crespo}, N., {D'Abrusco}, R., {et~al.} 2016, \apss,
  361, 337, \dodoi{10.1007/s10509-016-2926-6}

\bibitem[{{Nievas Rosillo} {et~al.}(2022){Nievas Rosillo}, {Dom{\'\i}nguez},
  {Chiaro}, {La Mura}, {Brill}, \& {Paliya}}]{Rosillo_2022}
{Nievas Rosillo}, M., {Dom{\'\i}nguez}, A., {Chiaro}, G., {et~al.} 2022,
  \mnras, 512, 137, \dodoi{10.1093/mnras/stac491}

\bibitem[{{Nolan} {et~al.}(2012){Nolan}, {Abdo}, {Ackermann}, {Ajello},
  {Allafort}, {Antolini}, {Atwood}, {Axelsson}, {Baldini}, {Ballet},
  {Barbiellini}, {Bastieri}, {Bechtol}, {Belfiore}, {Bellazzini}, {Berenji},
  {Bignami}, {Blandford}, {Bloom}, {Bonamente}, {Bonnell}, {Borgland},
  {Bottacini}, {Bouvier}, {Brandt}, {Bregeon}, {Brigida}, {Bruel}, {Buehler},
  {Burnett}, {Buson}, {Caliandro}, {Cameron}, {Campana}, {Ca{\~n}adas},
  {Cannon}, {Caraveo}, {Casandjian}, {Cavazzuti}, {Ceccanti}, {Cecchi},
  {{\c{C}}elik}, {Charles}, {Chekhtman}, {Cheung}, {Chiang}, {Chipaux},
  {Ciprini}, {Claus}, {Cohen-Tanugi}, {Cominsky}, {Conrad}, {Corbet}, {Cutini},
  {D'Ammando}, {Davis}, {de Angelis}, {DeCesar}, {DeKlotz}, {De Luca}, {den
  Hartog}, {de Palma}, {Dermer}, {Digel}, {Silva}, {Drell}, {Drlica-Wagner},
  {Dubois}, {Dumora}, {Enoto}, {Escande}, {Fabiani}, {Falletti}, {Favuzzi},
  {Fegan}, {Ferrara}, {Focke}, {Fortin}, {Frailis}, {Fukazawa}, {Funk},
  {Fusco}, {Gargano}, {Gasparrini}, {Gehrels}, {Germani}, {Giebels},
  {Giglietto}, {Giommi}, {Giordano}, {Giroletti}, {Glanzman}, {Godfrey},
  {Grenier}, {Grondin}, {Grove}, {Guillemot}, {Guiriec}, {Gustafsson},
  {Hadasch}, {Hanabata}, {Harding}, {Hayashida}, {Hays}, {Hill}, {Horan},
  {Hou}, {Hughes}, {Iafrate}, {Itoh}, {J{\'o}hannesson}, {Johnson}, {Johnson},
  {Johnson}, {Johnson}, {Kamae}, {Katagiri}, {Kataoka}, {Katsuta}, {Kawai},
  {Kerr}, {Kn{\"o}dlseder}, {Kocevski}, {Kuss}, {Lande}, {Landriu},
  {Latronico}, {Lemoine-Goumard}, {Lionetto}, {Llena Garde}, {Longo},
  {Loparco}, {Lott}, {Lovellette}, {Lubrano}, {Madejski}, {Marelli}, {Massaro},
  {Mazziotta}, {McConville}, {McEnery}, {Mehault}, {Michelson}, {Minuti},
  {Mitthumsiri}, {Mizuno}, {Moiseev}, {Mongelli}, {Monte}, {Monzani},
  {Morselli}, {Moskalenko}, {Murgia}, {Nakamori}, {Naumann-Godo}, {Norris},
  {Nuss}, {Nymark}, {Ohno}, {Ohsugi}, {Okumura}, {Omodei}, {Orlando}, {Ormes},
  {Ozaki}, {Paneque}, {Panetta}, {Parent}, {Perkins}, {Pesce-Rollins},
  {Pierbattista}, {Pinchera}, {Piron}, {Pivato}, {Porter}, {Racusin},
  {Rain{\`o}}, {Rando}, {Razzano}, {Razzaque}, {Reimer}, {Reimer}, {Reposeur},
  {Ritz}, {Rochester}, {Romani}, {Roth}, {Rousseau}, {Ryde}, {Sadrozinski},
  {Salvetti}, {Sanchez}, {Saz Parkinson}, {Sbarra}, {Scargle}, {Schalk},
  {Sgr{\`o}}, {Shaw}, {Shrader}, {Siskind}, {Smith}, {Spandre}, {Spinelli},
  {Stephens}, {Strickman}, {Suson}, {Tajima}, {Takahashi}, {Takahashi},
  {Tanaka}, {Thayer}, {Thayer}, {Thompson}, {Tibaldo}, {Tibolla}, {Tinebra},
  {Tinivella}, {Torres}, {Tosti}, {Troja}, {Uchiyama}, {Vandenbroucke}, {Van
  Etten}, {Van Klaveren}, {Vasileiou}, {Vianello}, {Vitale}, {Waite},
  {Wallace}, {Wang}, {Werner}, {Winer}, {Wood}, {Wood}, {Wood}, {Yang}, \&
  {Zimmer}}]{2FGL_2012}
{Nolan}, P.~L., {Abdo}, A.~A., {Ackermann}, M., {et~al.} 2012, \apjs, 199, 31,
  \dodoi{10.1088/0067-0049/199/2/31}

\bibitem[{Padovani(1992)}]{padovani92}
Padovani, P. 1992, Monthly Notices of the Royal Astronomical Society, 257, 404,
  \dodoi{10.1093/mnras/257.3.404}

\bibitem[{Padovani {et~al.}(2002)Padovani, Costamante, Ghisellini, Giommi, \&
  Perlman}]{Padovani2002}
Padovani, P., Costamante, L., Ghisellini, G., Giommi, P., \& Perlman, E. 2002,
  \apj, 581, 895, \dodoi{10.1086/344406}

\bibitem[{{Paggi} {et~al.}(2014){Paggi}, {Milisavljevic}, {Masetti},
  {Jim{\'e}nez-Bail{\'o}n}, {Chavushyan}, {D'Abrusco}, {Massaro}, {Giroletti},
  {Smith}, {Margutti}, {Tosti}, {Mart{\'{\i}}nez-Galarza},
  {Ot{\'{\i}}-Floranes}, {Landoni}, {Grindlay}, \& {Funk}}]{paggi14}
{Paggi}, A., {Milisavljevic}, D., {Masetti}, N., {et~al.} 2014, \aj, 147, 112,
  \dodoi{10.1088/0004-6256/147/5/112}

\bibitem[{{Paiano} {et~al.}(2017){Paiano}, {Falomo}, {Franceschini}, {Treves},
  \& {Scarpa}}]{paiano17}
{Paiano}, S., {Falomo}, R., {Franceschini}, A., {Treves}, A., \& {Scarpa}, R.
  2017, \apj, 851, 135, \dodoi{10.3847/1538-4357/aa9af4}

\bibitem[{{Pe{\~n}a-Herazo} {et~al.}(2019){Pe{\~n}a-Herazo}, {Massaro},
  {Chavushyan}, {Marchesini}, {Paggi}, {Landoni}, {Masetti}, {Ricci},
  {D'Abrusco}, {Milisavljevic}, {Jim{\'e}nez-Bail{\'o}n}, {La Franca}, {Smith},
  \& {Tosti}}]{Pena-Herazo_2019}
{Pe{\~n}a-Herazo}, H.~A., {Massaro}, F., {Chavushyan}, V., {et~al.} 2019,
  \apss, 364, 85, \dodoi{10.1007/s10509-019-3574-4}

\bibitem[{{Pe{\~n}a-Herazo} {et~al.}(2021){Pe{\~n}a-Herazo}, {Paggi},
  {Garc{\'\i}a-P{\'e}rez}, {Amaya-Almaz{\'a}n}, {Massaro}, {Ricci},
  {Chavushyan}, {Marchesini}, {Masetti}, {Landoni}, {D'Abrusco},
  {Milisavljevic}, {Jim{\'e}nez-Bail{\'o}n}, {Pati{\~n}o-{\'A}lvarez}, {La
  Franca}, {Smith}, \& {Tosti}}]{Pena-Herazo_2021}
{Pe{\~n}a-Herazo}, H.~A., {Paggi}, A., {Garc{\'\i}a-P{\'e}rez}, A., {et~al.}
  2021, \aj, 162, 177, \dodoi{10.3847/1538-3881/ac1da7}

\bibitem[{{Rajagopal} {et~al.}(2021){Rajagopal}, {Marchesi}, {Kaur},
  {Dom{\'\i}nguez}, {Silver}, \& {Ajello}}]{rajagopal2021}
{Rajagopal}, M., {Marchesi}, S., {Kaur}, A., {et~al.} 2021, \apjs, 254, 26,
  \dodoi{10.3847/1538-4365/abf656}

\bibitem[{{Ricci} {et~al.}(2015){Ricci}, {Massaro}, {Landoni}, {D'Abrusco},
  {Milisavljevic}, {Stern}, {Masetti}, {Paggi}, {Smith}, \& {Tosti}}]{ricci15}
{Ricci}, F., {Massaro}, F., {Landoni}, M., {et~al.} 2015, \aj, 149, 160,
  \dodoi{10.1088/0004-6256/149/5/160}

\bibitem[{{Sbarrato} {et~al.}(2012){Sbarrato}, {Ghisellini}, {Maraschi}, \&
  {Colpi}}]{sbarrato2012}
{Sbarrato}, T., {Ghisellini}, G., {Maraschi}, L., \& {Colpi}, M. 2012, \mnras,
  421, 1764, \dodoi{10.1111/j.1365-2966.2012.20442.x}

\bibitem[{Shakura \& Sunyaev(1973)}]{shakura1973black}
Shakura, N.~I., \& Sunyaev, R.~A. 1973, Astronomy and Astrophysics, 24, 337

\bibitem[{{Shaw} {et~al.}(2013){Shaw}, {Romani}, {Cotter}, {Healey},
  {Michelson}, {Readhead}, {Richards}, {Max-Moerbeck}, {King}, \&
  {Potter}}]{shaw13}
{Shaw}, M.~S., {Romani}, R.~W., {Cotter}, G., {et~al.} 2013, \apj, 764, 135,
  \dodoi{10.1088/0004-637X/764/2/135}

\bibitem[{Silver {et~al.}(2020)Silver, Marchesi, Marcotulli, Kaur, Rajagopal,
  \& Ajello}]{silver2020}
Silver, R., Marchesi, S., Marcotulli, L., {et~al.} 2020, \apj, 902, 23,
  \dodoi{10.3847/1538-4357/abb317}

\bibitem[{Tavecchio {et~al.}(2011)Tavecchio, Ghisellini, Bonnoli, \&
  Foschini}]{Tavecchio_2011}
Tavecchio, F., Ghisellini, G., Bonnoli, G., \& Foschini, L. 2011, Monthly
  Notices of the Royal Astronomical Society, 414, 3566,
  \dodoi{10.1111/j.1365-2966.2011.18657.x}

\bibitem[{{Taylor}(2005)}]{taylor05}
{Taylor}, M.~B. 2005, in Astronomical Society of the Pacific Conference Series,
  Vol. 347, Astronomical Data Analysis Software and Systems XIV, ed.
  P.~{Shopbell}, M.~{Britton}, \& R.~{Ebert}, 29

\bibitem[{{Urry} \& {Padovani}(1995)}]{urry95}
{Urry}, C.~M., \& {Padovani}, P. 1995, \pasp, 107, 803, \dodoi{10.1086/133630}

\end{thebibliography}

\end{document}